\documentclass[a4paper,11pt]{article}

\usepackage{jheppub2}
\usepackage{dsfont}
\usepackage{amsmath,amssymb,amscd,amsfonts,mathtools}
\usepackage{amsthm,thmtools}
\usepackage{xcolor}
\usepackage{tikz}
\usetikzlibrary{quantikz}
\theoremstyle{plain}

\usepackage[toc,page]{appendix}
\usepackage{amsthm}
\usepackage{epsfig}
\usepackage{epstopdf}
\usepackage{latexsym}
\usepackage{graphicx}
\usepackage{placeins}
\usepackage{floatrow}
\usepackage{subfig}
\usepackage{caption}
\usepackage{booktabs}
\usepackage{bbm}
\usepackage{color}
\usepackage{physics}
\usepackage{stackrel}
\usepackage{tensor}
\usepackage{tikz}
\usepackage{lipsum}
\usepackage{comment}

\usetikzlibrary{matrix}
\usetikzlibrary{decorations.markings,calc,shapes,decorations.pathmorphing,patterns,decorations.pathreplacing}
\usetikzlibrary{positioning}
\usepackage{hyperref}

\pdfoutput=1
\makeatletter
\def\@fpheader{\relax}
\makeatother

\newtheorem*{theorem*}{Theorem}

\newcommand{\beq}{\begin{equation}}
\newcommand{\eeq}{\end{equation}}

\DeclareMathSizes{10}{10}{7}{7}

\subheader{\begin{flushright}
\texttt{IFT-UAM/CSIC-25-30}\\
\end{flushright}}

\title{The landscape of complexity measures in 2D gravity}

\author[a]{Elena C\'aceres,\!}
\author[a,b]{Rafael Carrasco,\!}
\author[a]{Vaishnavi Patil,\!}
\author[b]{Juan F. Pedraza,\!}
\author[c]{Andrew Svesko}

\affiliation[a]{Theory Group, Weinberg Institute, Department of Physics, University of Texas at Austin,\\
2515 Speedway, Austin, Texas 78712, USA.}
\affiliation[b]{Instituto de Física Teórica UAM/CSIC, Calle Nicolás Cabrera 13-15, Madrid 28049, Spain}
\affiliation[c]{Department of Mathematics, King’s College London, Strand, London, WC2R 2LS, UK}

\emailAdd{elenac@utexas.edu}
\emailAdd{rafael.carrasco@ift.csic.es}
\emailAdd{vaishnavi.patil@utexas.edu}
\emailAdd{j.pedraza@csic.es}
\emailAdd{andrew.svesko@kcl.ac.uk}

\abstract{We investigate the broad landscape of holographic complexity measures for theories dual to two-dimensional (2D) dilaton gravity. Previous studies have largely focused on the complexity=volume and complexity=action proposals for holographic complexity. Here we systematically construct and analyze a wide class of generalized complexity functionals, focusing on codimension-one bulk observables. Two complementary approaches are presented: one inspired by dimensional reduction of codimension-one observables from higher-dimensional gravity, and another that adopts a purely 2D perspective. We verify the resulting observables exhibit hallmark features of complexity, such as linear growth at late times and the switchback effect. We further offer heuristic interpretations of the role of multiple extremal surfaces when they appear.  Finally, we comment on the bulk-to-boundary dictionary via the covariant Peierls bracket in 2D gravity.
Our work lays the groundwork for a richer understanding of quantum complexity in low-dimensional holographic dualities.} 

\begin{document}

\maketitle

\section{Introduction} \label{sec:intro}

Characteristic traits of a generic black hole are its event horizon and the singularity deep within. Holographic duality
is an established framework capable of investigating each of these features using information theoretic measures. 
For example, the Ryu-Takayanagi prescription \cite{Ryu:2006bv} shows the area of minimal (or extremal \cite{Hubeny:2007xt}) codimension-2 surfaces in asymptotically anti-de Sitter (AdS) spacetimes, e.g., black hole horizons, is microscopically captured by the entanglement entropy between subregions in a dual conformal field theory (CFT) confined to the boundary of `bulk' AdS (assuming AdS/CFT duality). Meanwhile, the late-time growth of the Einstein-Rosen bridge inside eternal AdS black holes is dual to complexity growth of holographic operators \cite{Susskind:2014rva,Susskind:2014moa}, promising a way to microscopically probe a black hole interior, including, possibly, the singularity (see \cite{Baiguera:2025dkc} for a review).

For ordinary quantum mechanical systems, computational complexity quantifies the smallest number of unitary operators, or gates, needed to obtain a particular target state from a given reference state, within a specified margin of error. A precise definition of complexity in field theories remains an open area of inquiry (see, e.g., \cite{Chapman:2017rqy,Jefferson:2017sdb,Caputa:2017urj,Caputa:2017yrh,Caputa:2018kdj,Chapman:2018hou,Hackl:2018ptj,Khan:2018rzm,Camargo:2019isp, Caceres:2019pgf,Flory:2020eot,Flory:2020dja,Chagnet:2021uvi,Chandra:2022pgl}). Still, it is natural to hypothesize  gravitational observables dual to complexity for a holographic CFT. A guiding principle for characterizing quantum complexity of holographic field theories, i.e., `holographic complexity', is that the complexity for quantum systems display two essential behaviors: (i) linear growth of complexity of the thermofield double state at late (boundary) times, and (ii) the switchback effect. For bulk observables, the former 
is attributed to the growth of the Einstein-Rosen bridge connecting a double-sided AdS black hole at a rate characterized by the mass and other thermodynamic potentials of the black hole.
The latter describes the time delay in the response of a bulk observable to CFT state perturbations in the far past, gravitationally represented as shockwaves \cite{Stanford:2014jda}.  It turns out there is an infinite class of equally viable candidate observables which capture the behaviors (i) and (ii), cheekily dubbed the `complexity=anything' (CAny) proposal \cite{Belin:2021bga,Belin:2022xmt} (see also \cite{Myers:2024vve}).

Concretely, there are two classes of observables: codimension-1 observables \cite{Belin:2021bga} and codimension-0 observables \cite{Belin:2022xmt}. For the first type, let $\Sigma$ be a codimension-1 bulk hypersurface in $(d+1)$-dimensional asymptotically AdS spacetime, anchored on a boundary Cauchy slice $\sigma_{\text{CFT}}$, such that $\partial\Sigma=\sigma_{\text{CFT}}$. Similar to the Ryu-Takayanagi prescription for holographic entanglement entropy, the holographic complexity $\mathcal{C}_{\text{gen}}$ is characterized by a two step process: (1) a maximization prescription to select the maximal bulk hypersurface among all possible spacelike surfaces with boundary fixed along $\sigma_{\text{CFT}}$, and (2) the evaluation of a specific diffeomorphism invariant quantity of said bulk maximal surface, 
\beq \mathcal{C}_{\text{gen}}(\sigma_{\text{CFT}})=\underset{\partial\Sigma=\sigma_{\text{CFT}}}{\text{max}}\left[\frac{1}{G_{\text{N}}L}\int_{\Sigma}d^{d}y\sqrt{h}F(g_{\mu\nu},X^\mu)\right]\;.\label{eq:canycod1intro}\eeq
Here $G_{\text{N}}$ denotes Newton's constant, $L$ is AdS$_{d+1}$ length scale, $h_{ij}$ is the induced metric on the bulk hypersurface $\Sigma$ (the integral sums over all possible surfaces), and $F$ is a scalar function of the bulk metric $g_{\mu\nu}$, its derivatives, and on the embedding functions $X^{\mu}_{\pm}(y^{a})$ of $\Sigma$.\footnote{A more general version of (\ref{eq:canycod1intro}) exists where there is a second scalar function used in the maximization prescription to determine $\Sigma$, independent of $F$---see \cite{Belin:2021bga}. Further, here we make the simplifying assumption that $F$ is independent of combinations of extrinsic curvature associated with $\Sigma$.} The simplest example occurs when $F=1$, such that the integral gives the volume of the bulk maximal Cauchy slice, recovering the first proposal for holographic complexity, `complexity=volume' (CV) \cite{Susskind:2014rva,Susskind:2014jwa,Stanford:2014jda}. More generally, for a large class of $F$, the observables exhibit both late time linear growth and the switchback effect, thus display the essential properties of complexity.
Notably, the definition (\ref{eq:canycod1intro}) is agnostic to the number of spacetime dimensions and theory of gravity exhibiting diffeomorphism invariance.\footnote{Generalized codimension-1 functionals appeared before \cite{Alishahiha:2015rta,Bueno:2016gnv,Hernandez:2020nem}, where the volume functional is corrected with higher-derivative contributions.} The large family of equally valid functionals is perhaps not surprising given that conventional computational complexity is ambiguous, e.g., choosing the set of unitary gates to perform the operations.

The family of codimension-0 observables are characterized as follows \cite{Belin:2022xmt}. Let $\mathcal{M}$ be a bulk codimension-0 region with past and future boundaries $\Sigma_{\pm}$ such that $\partial\mathcal{M}=\Sigma_{+}\cup\Sigma_{-}$, anchored at a boundary timeslice $\sigma_{\text{CFT}}$, $\partial\Sigma_{\pm}=\sigma_{\text{CFT}}$. Then another candidate for holographic complexity follows a similar two-step extremization procedure,
\beq 
\begin{split}
\mathcal{C}_{\text{gen}}(\sigma_{\text{CFT}})&=\underset{\partial\Sigma_{\pm}=\sigma_{\text{CFT}}}{\text{max}} \biggr[\frac{1}{G_{\text{N}}L^{2}}\int_{\mathcal{M}_{G,F_{\pm}}}d^{d+1}x\sqrt{-g}G(g_{\mu\nu})\\
&+\frac{1}{G_{\text{N}}L}\int_{\Sigma_{+}}d^{d}y\sqrt{h}F_{+}(g_{\mu\nu};X^{\mu}_{+})+\frac{1}{G_{\text{N}}L}\int_{\Sigma_{-}}d^{d}y\sqrt{h}F_{-}(g_{\mu\nu};X^{\mu}_{-})\biggr]\;,
\end{split}
\label{eq:canycod0intro}\eeq
where $G$ and $F_{\pm}$ are in principle independent scalar functionals. Now the maximization procedure independently varies over the embeddings
to extremize the two codimension-1 boundary integrals plus the bulk integral which is evaluated over the codimension-0 region $\mathcal{M}_{G,F_{\pm}}$. The functional (\ref{eq:canycod0intro}) includes the codimension-1 observables (\ref{eq:canycod1intro}) as a special case.  Further, the codimension-0 observables generalize the previous `complexity=action' (CA) conjecture \cite{Brown:2015bva,Brown:2015lvg,Fan:2018wnv}, where complexity is equal to the gravitational action evaluated on the Wheeler-De Witt (WDW) patch,\footnote{Formally, the WDW patch is the domain of dependence of any bulk Cauchy surface that asymptotically approaches the boundary time slice.} and `complexity=spacetime volume' (CV2.0) proposal \cite{Couch:2016exn}, where one evaluates the spacetime volume of the WDW patch. Both proposals (\ref{eq:canycod1intro}) and (\ref{eq:canycod0intro}) were later refined in \cite{Jorstad:2023kmq} to address certain ambiguities we comment on below, and to systematically probe the singularity structure of AdS black holes (see also \cite{Arean:2024pzo,Caceres:2024edr}). 

In this article, we will explore the set of holographic complexity functionals that are allowed in two-dimensional (2D) theories of dilaton gravity. 
Such low-dimensional models generically include a scalar field non-minimally coupled to the metric and therefore do not fall into the same universality class as their higher-dimensional counterparts. However, they provide analytically solvable setups that capture universal features of higher-dimensional black holes, such as their horizon thermodynamics and quantum chaotic behavior.
Pertinently, these 2D models serve as testbeds for holographic complexity proposals, facilitating comparisons with an appropriate dual quantum mechanical theory.  For example, Jackiw-Teitelboim (JT) gravity \cite{Jackiw:1984je,Teitelboim:1983ux} captures deviations away from extremality of charged or rotating black holes \cite{Achucarro:1993fd,Fabbri:2000xh,Nayak:2018qej,Sachdev:2019bjn,Castro:2018ffi,Moitra:2019bub,Castro:2021fhc} and  holographically reproduces the near-conformal dynamics of the Sachdev-Ye-Kitaev (SYK) \cite{Sachdev:1992fk,Kit_SYK} model of coupled fermions \cite{Maldacena:2016upp}, providing a concrete setting to explore various proposals of holographic complexity \cite{Brown:2018bms}. In fact, aspects of CV and its connection to  Krylov complexity in the SYK model  have already been studied  in the literature, exhibiting, e.g., 
the same exponential-to-linear growth behavior \cite{Jian:2020qpp}.\footnote{In \cite{Rabinovici:2023yex,Ambrosini:2024sre,Heller:2024ldz} the connection between CV in JT and some versions of the SYK model was further analyzed.}

Another advantage of working in lower-dimensional gravity is 
that one can exactly incorporate non-perturbative and semi-classical quantum effects, (which would otherwise be prohibitively challenging in higher dimensions).
Indeed, the inclusion of non-perturbative effects is necessary to obtain the expected late-time saturation of complexity \cite{Iliesiu:2021ari, Balasubramanian:2024lqk, Gautason:2025ryg, Miyaji:2025yvm}. Similarly, integrable irrelevant deformations ($T\bar{T}$ deformations) were found to modify the energy spectrum non-trivially, making the ramp of the Spectral Form Factor to 
rise faster than in the undeformed theory \cite{Bhattacharyya:2025gvd, Bhattacharyya:2023gvg}. Further, using semi-classical JT gravity, it was found how to generalize CV complexity to include corrections from bulk quantum fields \cite{Carrasco:2023fcj}. 

Thus far, proposals for holographic complexity explored in 2D gravity have largely focused on CV and CA, while a comprehensive exploration of `complexity=anything' for 2D models of gravity remains to be developed. We fill this gap by exploring the landscape of complexity functionals for a general class of 2D dilaton-gravity models, focusing on codimension-1 observables, following earlier studies in higher-dimensional gravity theories. Our work thus aims to benchmark holographic complexity in AdS$_2$/CFT$_1$.

Our paper is organized as follows. In Section \ref{sec:CA4D},  we succinctly review the generic family of codimension-1 observables for four-dimensional charged AdS black holes. In this context, linear growth at late times of the observable is known to be a consequence of an effective one-dimensional potential having a local maxima. Non-extremal charged black holes, however, generally give rise to potentials with multiple maxima. We provide two novel interpretations of the existence of the subleading maxima: (i) each maxima serves as a distinct, locally optimal way to prepare the same final state, and (ii) all local maxima additively contribute to the total complexity. 

In Section \ref{sec:CA2D} we develop the proposal for codimension-1 generalized measures of complexity for a wide class of two-dimensional dilaton gravity theories. Our proposal is in part motivated by the specific class of observables found from a spherical dimensional reduction of the codimension-1 observables used to characterize holographic complexity for higher-dimensional charged black holes. We further provide a more general proposal 
 of codimension-1 observables (cf. (\ref{eq:complexfam2d}))
\begin{equation}
    \mathcal{C}_{\textup{gen}}=\frac{1}{G_2 L}\int_{\Sigma_{F_2}} dy \sqrt{h}F_1(g_{\mu\nu},\Phi, X^{\mu}),
\label{eq:complexfam2dintro}\end{equation}
for a broad class of diffeomorphism invariant functions $F_1$ and $F_{2}$ of the metric and dilaton $\Phi$. Evidence for our proposal stems from the fact these observables exhibit late linear time growth and switchback effect.

We comment on the bulk-to-boundary
dictionary of generalized complexity observables in Section \ref{sec:dictionary} using the language of covariant phase space. In particular, we argue  complexity=anything provides a dictionary: once a specific bulk functional is chosen, the boundary field theory must have a matching complexity functional, obeying the first law of complexity for that specific bulk functional. Further,  we argue that holographic complexity has a robust, though innately scheme dependent gravitational dual.

We conclude with an outlook in Section \ref{sec:disc}, where we provide an outlook for future work. Appendix \ref{app:dimredx} presents explicit formulae for the dimensional reduction of higher-dimensional curvature invariants. In Appendix \ref{app:CVforJT}, we adapt the covariant Peierls bracket to the case of complexity=volume for JT gravity, to lay the groundwork for a holographic dictionary of complexity measures in 2D.


\section{Complexity=anything for charged AdS black holes} \label{sec:CA4D}

Here we review, filling in some gaps along the way, the complexity=anything proposal applied to charged-AdS$_{4}$ black holes. These backgrounds will prove relevant when we construct the analog of CAny observables for two-dimensional dilaton gravity theories. 

\subsection*{Charged AdS black holes}

Here we focus on the spherically symmetric Reissner-N{\"o}rdstrom metric in AdS$_{4}$ with cosmological constant $\Lambda=-3/L^{2}$. In static coordinates the line element has the form
\beq ds^{2}= -f(r)dt^2 + f^{-1}(r)dr^{2} + r^2 d\Omega_2^{2}\;,\label{eq:RNbhmet}\eeq
with blackening factor
\beq f(r) = 1 - \frac{2 MG_{\text{N}}}{r} + \frac{Q^2}{r^2} + \frac{r^2}{L^2}\;.\eeq
Here $M$ and $Q$ are parameters proportional to the mass and electric charge of the black hole. For non-extremal black holes, the blackening factor has two positive roots, $f(r_{\pm})=0$, satisfying $r_{-}<r_{+}$, where $r_{+}$ ($r_{-}$) denotes the outer (inner) event horizon. In the extremal limit, where $r_{+}=r_{-}$, the surface gravity 
vanishes, and the near-horizon extremal geometry has the product form AdS$_{2}\times S^{2}$.

The metric (\ref{eq:RNbhmet}) is a solution to Einstein-Maxwell-AdS$_{4}$ gravity theory,
\beq\label{eq:RNaction} I_{\text{EM}}=\frac{1}{16\pi G_{\text{N}}}\int_{\mathcal{M}}d^{4}x\sqrt{-g}\left(R-2\Lambda-\frac{1}{4\mu_{0}}F_{\mu\nu}^{2}\right)\;,\eeq
for electromagnetic coupling $\mu_{0}$,\footnote{The solution (\ref{eq:RNbhmet}) is in units $\mu_{0}=1$.} obeying the Einstein-Maxwell field equations
\beq G_{\mu\nu}+\Lambda g_{\mu\nu}=8\pi G_{\text{N}}T_{\mu\nu}\;,\quad T_{\mu\nu}\equiv-\frac{2}{\sqrt{-g}}\frac{\delta I_{\text{Max}}}{\delta g^{\mu\nu}}=\frac{1}{16\pi G_{\text{N}}\mu_{0}}\left(F^{\rho}_{\phantom{\rho}\mu}F_{\rho\nu}-\frac{1}{4}g_{\mu\nu}F^{2}_{\rho\sigma}\right)\;,\eeq
\beq \nabla_{\mu}F^{\mu\nu}=0\;,\eeq
with $U(1)$ gauge field $A_{\mu}$
\beq 
\begin{split}
 A=A_{\mu}dx^{\mu}=Q\left(\frac{1}{r_{+}}-\frac{1}{r}\right)dt\;,  
\end{split}
\label{eq:AfieldKN}\eeq
and Maxwell field strength tensor $F_{\mu\nu}=\partial_{\mu}A_{\nu}-\partial_{\nu}A_{\mu}$.

\begin{figure}[t!]
    \centering
    \includegraphics[width=0.5\linewidth]{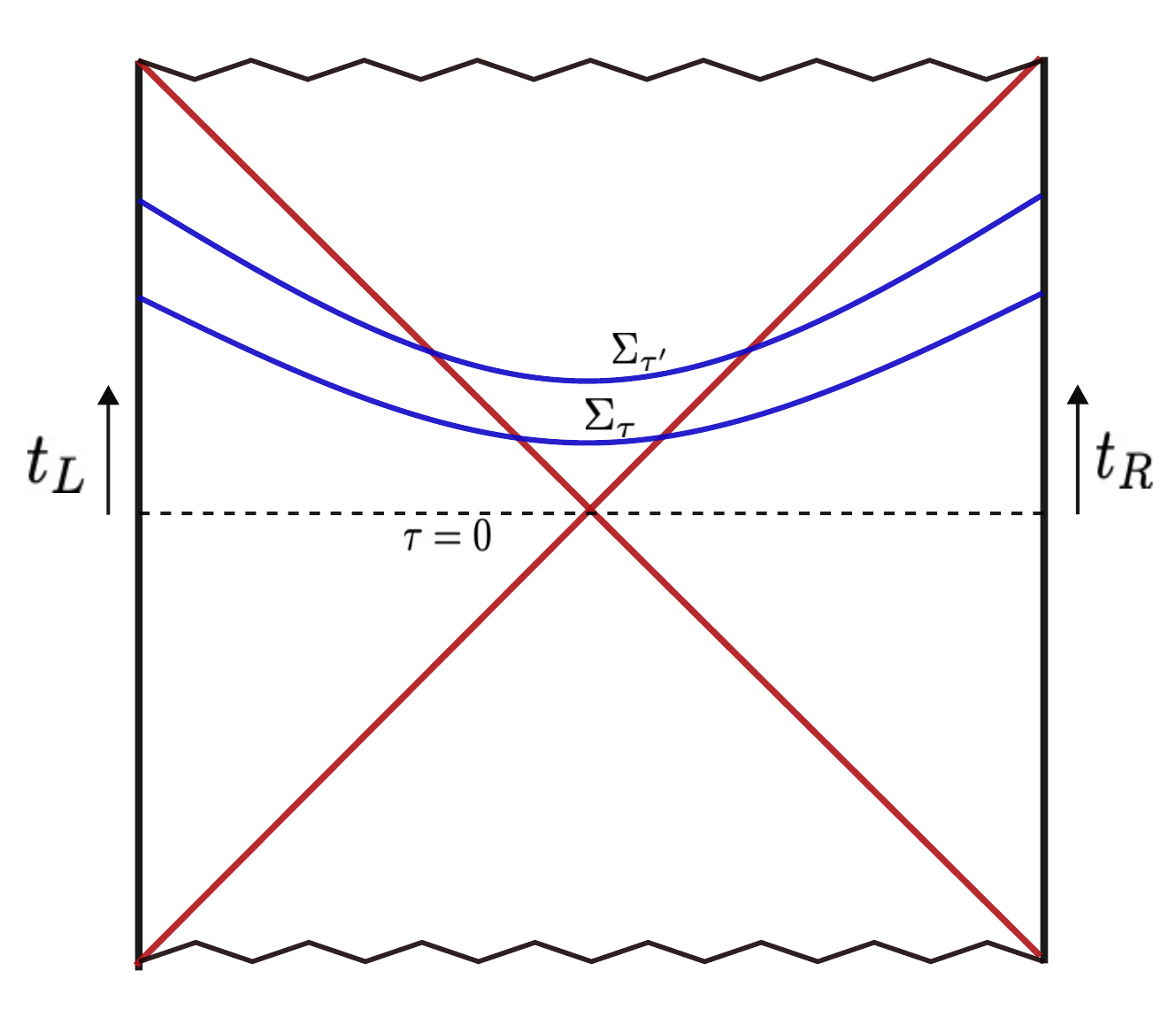}
    \caption{Thermofield double state dual to a (neutral) two sided eternal black hole. The diagonal lines (red) denote the black hole event horizon. The extremal codimension-1 hypersurface (in blue) $\Sigma_{\tau}$ time evolves into the extremal surface $\Sigma_{\tau'}$, the evolution of which has a dual description in terms of the TFD evolution.}
    \label{fig:TFD cartoon}
\end{figure}

To evaluate complexity observables, it is often useful to work in (ingoing) Eddington-Finkelstein coordinates so as to cover both the exterior and interior horizon geometry. The line element is 
\beq ds^{2}=-f(r)dv^{2}+2dvdr+r^{2}d\Omega_2^{2}\;,\label{eq:EFcoord}\eeq
for infalling coordinate $v=t+r_{\ast}(r)$ and tortoise coordinate $r_{\ast}(r)=-\int_{r}^{\infty}dr'/f(r')$. Holographically, the global two-sided black hole is dual to two decoupled CFTs living on the left/right sided timelike AdS$_{4}$ boundaries, entangled in the (charged) thermofield double (TFD) state
\beq |\psi_{\text{TFD}}(t_{L},t_{R})\rangle=\sum_{n,\alpha}e^{-\beta(E_{n}-\mu Q_{\alpha})/2-iE_{n}(t_{L}+t_{R})}|E_{n},-Q_{\alpha}\rangle_{L}\otimes |E_{n},Q_{\alpha}\rangle_{R}\;.\label{eq:TFDstate}\eeq
Here $t_{L,R}$ denote left/right boundary times, $Q_{\alpha}$ the charge, $\mu$ a chemical potential, inverse temperature $\beta$, and $|E_{n},Q_\alpha\rangle_{L,R}$ the left/right charged energy eigenstate. In other words, the spacetime (\ref{eq:EFcoord}) characterizes the time evolution of the dual TFD state  living at a timeslice with $\tau=(t_{L}+t_{R})$ (for simplicity often $t_{L}=t_{R}$, as we will adopt here), see Figure \ref{fig:TFD cartoon} for an illustration. 

\subsection{Generalized volumes and codimension-one observables}

As briefly outlined above, the essential prescription evaluating the (generalized) holographic complexity $\mathcal{C}_{\text{gen}}$ of a boundary Cauchy slice $\sigma$ using codimension-1 observables (\ref{eq:canycod1intro}) is the following \cite{Belin:2021bga} (see also the summary in \cite{Belin:2022xmt,Jorstad:2023kmq}). First find a codimension-1 spatial slice $\Sigma$ of the bulk AdS$_{d+1}$ spacetime satisfying $\partial\Sigma=\sigma_{\text{CFT}}$ that extremizes an arbitrary scalar functional $F_{2}$ of the background metric $g_{\mu\nu}$, curvature invariants, and the embedding function $X^{\mu}(y^{a})$ of $\Sigma$. That is, 
\beq
\delta_X \left( \int_\Sigma d^{d}y\sqrt{h}F_2(g_{\mu\nu},X^\mu) \right) = 0\;.\label{eq:extremizationcodim1}\eeq
Extremization (\ref{eq:extremizationcodim1}) uncovers an extremal codimension-1 hypersurface $\Sigma_{F_{2}}$ (of family thereof) which is then used to evaluate the complexity
\beq 
\mathcal{C}_{\text{gen}} = \frac{1}{G_N L} \int_{\Sigma_{F_2}} d^{d}y\sqrt{h}F_1(g_{\mu\nu},X^\mu)\;,\label{eq:cgencod1}\eeq
for another scalar function $F_{1}$. Moving forward, we take the simplifying case when $F_{1}=F_{2}\equiv F$, i.e., the same scalar function is used to determine the extremal slice and evaluate the observable, leading to the prescription (\ref{eq:canycod1intro}). 

A reason the family $\mathcal{C}_{\text{gen}}$ (\ref{eq:cgencod1}) is thought to be dual to holographic complexity is because for some choices of $F_{1,2}$ they exhibit linear growth at late times, coinciding with the behavior of the computational (circuit) complexity of thermofield double states. This was explicitly shown for planar-AdS black holes in \cite{Belin:2021bga} and charged-AdS black holes in \cite{Jorstad:2023kmq} (see also \cite{Wang:2023eep,Jiang:2023jti}). Let us revisit linear growth for RN-AdS black holes, filling in some gaps along the way. 

Let $\rho$ denote a radial coordinate on the worldvolume of $\Sigma$ such that the codimension-1 spacelike hypersurfaces are parametrized as $(v(\rho),r(\rho),\theta,\phi)$, where $v$ is the infalling coordinate in metric (\ref{eq:EFcoord}) and $(\theta,\phi)$ denote the angular coordinates. With this parametrization, the family of codimension-1 observables (\ref{eq:cgencod1}) is easily cast as
\beq \mathcal{C}_{\text{gen}}=\frac{4\pi L}{G_{\text{N}}}\int_{\Sigma}d\rho\left(\frac{r}{L}\right)^{2}\sqrt{-f(r)\dot{v}^{2}+2\dot{v}\dot{r}}a(r)\equiv \frac{4\pi L}{G_{\text{N}}}\int_{\Sigma}d\rho \mathcal{L}_{\text{gen}}\;,\label{eq:codim1obser4DRN}\eeq
where $\dot{v}\equiv\partial_{\rho}v$ (and similarly for $\dot{r}$), and $a(r)$ denotes the evaluation of the scalar functional $F$ on $\Sigma$, which is purely a function of $r$ for such geometries.  Finding extremal surfaces as stipulated in (\ref{eq:extremizationcodim1}) then amounts to solving the classical equations of motion for Lagrangian $\mathcal{L}_{\text{gen}}$. 

Since $\mathcal{L}_{\text{gen}}$ is independent of $v$ ($\partial_{v}\mathcal{L}_{\text{gen}}=0$), it follows from the Euler-Lagrange equations that the system has a conserved momentum
\beq P_{v}\equiv \frac{\partial \mathcal{L}_{\text{gen}}}{\partial \dot{v}}=a(r)\left(\frac{r}{L}\right)^{2}\frac{(\dot{r}-f(r)\dot{v})}{\sqrt{-f\dot{v}^{2}+2\dot{v}\dot{r}}}\;.\eeq
It is convenient to gauge fix such that $a(r)r^{2}/L^{2}=\sqrt{-f\dot{v}^{2}+2\dot{v}\dot{r}}$, and the conserved momentum conjugate to $v$ becomes
\beq P_{v}=\dot{r}-f(r)\dot{v}\;.\label{eq:consmomRN}\eeq
As such, the problem of finding the extremal slice $\Sigma$ has been reduced to solving the equation of motion of a classical non-relativistic particle in a potential $U(r)$ \cite{Belin:2021bga}\footnote{This directly follows from combining the gauge fixing condition with the momentum (\ref{eq:consmomRN}). In the process it is also easy to uncover two extremality conditions $\dot{r}=\pm\sqrt{P_{v}^{2}+fa^{2}r^{4}}$ and $\dot{t}=\dot{v}-\frac{\dot{r}}{f}=-P_{v}\dot{r}/(f\sqrt{P_{v}^{2}+fa^{2}r^{4}})$.\label{fn:extconds}}
\beq \dot{r}^{2}+U(r)=P_{v}^{2}\;,\quad U=-f(r)a^2(r) \left( \frac{r}{L} \right)^{4}\;.\label{eq:extremPot}\eeq
Evidently, the effective potential vanishes when $r=r_{\pm}$, unless $a^{2}(r_{\pm})$ diverges faster.\footnote{If $a(r)$ diverged faster one would find $U(r)$ is such that the extremal surface would never penetrate the interior of the black hole.  Thus, every chosen function $a(r)$ must remain finite at the horizon or have a sufficiently mild divergence to guarantee the bulk observable is physically well-defined.}
For a given momentum $P_{v}$, the classical particle bounces off the potential at a turnaround point $r=r_{\text{min}}$, where $\dot{r}^{2}=0$ and the value of the potential is the maximum for that momentum $U(r_{\textup{min}})=P_v^2$. Correspondingly, for trajectories that begin and end at the asymptotic AdS boundaries, $r_{\text{min}}$ denotes the minimal radius reached by the extremal surface $\Sigma_{\text{ext}}$. Further, local maxima of $U(r)$, denoted $r=r_{f}$, correspond to a critical value of $r_{\text{min}}$, where $P_{v}$ is tuned to particular value such that $U'(r_{f})=0$ and $U''(r_{f})\leq0$.

We will test the linear growth for three different non-trivial curvature scalars.\footnote{The Ricci scalar is trivial, $R=-12/L^2$, so it reproduces standard complexity=volume.} Specifically, for the non-extremal RN-AdS$_{4}$ metric (\ref{eq:RNbhmet}), we consider (with $G_{\text{N}}=1$)
\beq
\begin{split}
&R_{\mu\nu}R^{\mu\nu} = \frac{36}{L^4} + \frac{4 Q^4}{r^8}\;,\\
&R_{\mu\nu\rho\sigma}R^{\mu\nu\rho\sigma} = 8 \left(\frac{3}{L^4}+\frac{6 M^2 r^2-12 M Q^2 r+7 Q^4}{r^8}\right)\;,\\
&C_{\mu\nu\rho\sigma}C^{\mu\nu\rho\sigma} = 48\frac{(Q^2-Mr)^2}{r^8}\;,
\end{split}
\label{eq:scalarcurvs}\eeq
where $C_{\mu\nu\rho\sigma}$ is the Weyl curvature tensor. For each scalar above, we may express the functional $a(r)$ appearing in the potential (\ref{eq:extremPot}) as
\begin{equation}
    a(r) = 1 + \lambda L^{4} (\text{scalar})
\end{equation}
for real constant $\lambda$, where $\lambda$ controls deviations away from the standard complexity=volume observables. 
We then scan the parameter range of $\lambda$ and $Q$ to determine where the corresponding potential $U(r)$ has local maxima. In the case of RN black holes, we impose the additional constraint that the maxima should lie between the two horizons, $r_{-}<r_{f}<r_{+}$. This is because the black hole solution becomes unstable inside the inner (Cauchy) horizon. 

\subsection*{Linear growth}

The infinite family of codimension-1 observables feature linear growth at late (boundary) times, an essential trait of complexity of thermofield double states. To see this, recall the codimension-1 observable (\ref{eq:codim1obser4DRN}). Using the gauge fixing $a(r)r^{2}/L^{2}=\sqrt{-f\dot{v}^{2}+2\dot{v}\dot{r}}$ and the effective extremization equation (\ref{eq:extremPot}) with conserved momentum $P_{v}$ (\ref{eq:consmomRN}) yields
\beq \mathcal{C}_{\text{gen}}=\frac{4\pi}{G_{\text{N}}L}\int_{r_{\text{min}}}^{\infty}dr\left(\frac{d\rho}{dr}\right)r^{4}a(r)^{2}=\frac{4\pi}{G_{\text{N}}L}\int_{r_{\text{min}}}^{\infty}dr\frac{r^{4}a(r)^{2}}{\sqrt{P_{v}^{2}-U(r)}}\;,\label{eq:Cgenworked}\eeq
for turning point $r_{\text{min}}$. Taking the derivative of the generalized complexity (\ref{eq:Cgenworked}) with respect to boundary time $\tau\equiv(t_{L}+t_{R})=2t_{R}$ (taking $t_{L}=t_{R}$ in TFD state (\ref{eq:TFDstate}))\footnote{This follows from advanced time $v=t+r^{\ast}$, i.e., $t_{R}+r^{\ast}(\infty)-r^{\ast}(r_{\text{min}})=\int dv \dot{v}=\int_{r_{\text{min}}}^{\infty}dr\frac{\dot{v}}{\dot{r}}=\int_{r_{\text{min}}}^{\infty}dr\frac{1}{f}\left(1-\frac{P_{v}}{\sqrt{P_{v}^{2}+fa^{2}r^{4}}}\right)$,
where the last equality follows from extremality conditions in Footnote \ref{fn:extconds}.}
\beq \tau=-2\int_{r_{\text{min}}}^{\infty}dr\frac{P_{v}}{f\sqrt{P_{v}^{2}-U(r)}}\;,\label{eq:bdrytime}\eeq
gives (see \cite{Belin:2021bga} for details)
\beq \frac{d\mathcal{C}_{\text{gen}}}{d\tau}=\frac{4\pi}{G_{\text{N}}L}P_{v}(\tau)\biggr|_{\partial\Sigma(\tau)}\;.\label{eq:timederivacomp}\eeq
Thus, the rate of change of the complexity evaluated at the extremal surface $\Sigma$ is given by the conjugate momentum evaluated at the endpoint $\partial\Sigma(\tau)$. Linear growth would then amount to $P_{v}|_{\partial\Sigma}$ being effectively constant. 

The condition of late time linear growth relies on the existence of at least one local maximum behind the horizon since, this fact implies the existence of an extremal surface that hits the boundary at infinity.
To wit, notice that the integrand for the boundary time $\tau$ (\ref{eq:bdrytime}) diverges\footnote{Due to the blackening factor $f(r)$, the integral (\ref{eq:bdrytime}) also diverges for values of $r$ near the horizon. This singular behavior is avoided by defining the integral by its Cauchy principal value associated with this singularity. That is, $\lim_{r\rightarrow r_{\pm}}\frac{P_{v}}{f\sqrt{P_{v}^{2}-U}}\sim \frac{r_{\pm}^{2}}{(r_{\pm}-r_{\mp})(r-r_{\pm})}$ for $r_{+}\neq r_{-}$. While the integrand itself diverges, its integral can be evaluated and is finite.}  when $r$ is near $r_{\text{min}}$ and $U(r)\approx P_{v}^{2}+U'(r_{\text{min}})(r-r_{\text{min}})$. This yields an integrable singularity,  such that the boundary time is finite, \emph{unless} the effective potential has a local maximum. Let $r_{f}<r_{+}$ be the critical value of $r_{\text{min}}$ such that $U(r)\approx P_{\infty}^{2}+\frac{1}{2}U''(r_{f})(r-r_{f})^{2}$ (with $U''(r_{f})\leq0$), where $P_{\infty}$ is the value of $P_{v}$ tuned such that $U'(r_{f})\to0$. Tuning the momentum as such, the singularity is no longer integrable and the boundary time diverges to positive infinity. Consequently, the $\tau\to\infty$ limit of the time derivative of generalized complexity (\ref{eq:timederivacomp}) yields a constant proportional to $P_{\infty}$. In terms of the effective potential, 
\beq \lim_{\tau\to\infty} \frac{d\mathcal{C}_{\text{gen}}}{d\tau}=\frac{4\pi}{G_{\text{N}}L}\sqrt{U(r_{f})}\;,\eeq
where we used $U(r_{f})=P_{\infty}^{2}$. Since the effective potential (\ref{eq:extremPot}) depends on the scalar function $a(r)$, the rate of linear growth will likewise depend on scalar functional $F$.\footnote{The blackening factor also features in the potential $U$, however, one can be agnostic to its particular form; all that is used is $f(r)$ has at least one real, positive root.}

\subsection*{Switchback Effect}

Briefly, let us summarize the second property that candidates for complexity are expected to satisfy, the switchback effect \cite{Stanford:2014jda}.  The switchback effect is described as follows. Perturb an initial thermofield double state by evolving with left/right Hamiltonians $H_{L,R}$ plus $n$ insertions of local operators at different points in time (out of order),
\begin{equation}
    \ket{\Psi(t_L,t_R)}=e^{-iH_Rt_R}e^{-iH_Lt_L}W_L(t_n)W_L(t_{n-1})...W_L(t_1)\ket{\Psi_{\text{TFD}}(0)}\;,
\label{eq:pertstateswitchback}\end{equation}
where times $t_1,t_2,...,t_n$ occur in alternating order, visualized as ``switchbacks'', and $W_{L}$ is some local operator acting at these points.
The forward and backward time evolution near each switchback creates partial cancellation of the operator insertions characterized by a specific timescale $t_*$ which is much less than the timescales of evolution between two insertions. 
The complexity of the state is captured by the total time evolution, i.e., the sum of the individual times between two switchbacks minus the partial cancellation $t_*$ \cite{Stanford:2014jda}, 
\begin{equation}
   \mathcal{C}\sim |t_1+t_R|+|t_2-t_1|+...+|t_L-t_n|-2nt_*\;.
\label{eq:switchbackeffectC}\end{equation}

Holographically, the perturbed state (\ref{eq:pertstateswitchback}) is characterized by a wormhole geometry with alternating left and right shockwaves. For strong shocks, the shockwave geometry obtained by gluing together multiple black hole geometries along their horizons with shifts along the null directions $u$ or $v$ (of Kruskal coordinates) \cite{Shenker:2013yza}. For the complexity=volume conjecture, the switchback effect (\ref{eq:switchbackeffectC}) manifests itself through two essential properties of extremal volume: the extremal volume is additive in the shockwave geometry for each shock, and each extremal volume segment exhibits late time linear growth.\footnote{The maximal volume slice is found by extremizing the additive volume including each segment, which then recovers the behavior (\ref{eq:switchbackeffectC}).}

The bulk codimension-1 observables characterizing $\mathcal{C}_{\text{gen}}$ exhibit both of the properties enjoyed by the extremal volume \cite{Belin:2021bga}. That the codimension-1 observables are additive is in part guaranteed by the diffeomorphism invariance of the functional $F_{1}$. Further, the general observables continue to exhibit late time linear growth, proportional to the effective potential $U(r_{f})$
on the final slice in the switchback geometry. Combined, the family of codimension-1 observables exhibit the switchback effect.

\subsection*{Potentials with multiple maxima}

\begin{figure}[t!]
        \subfloat[$R^{2}$]{%
            \includegraphics[width=.4\linewidth]{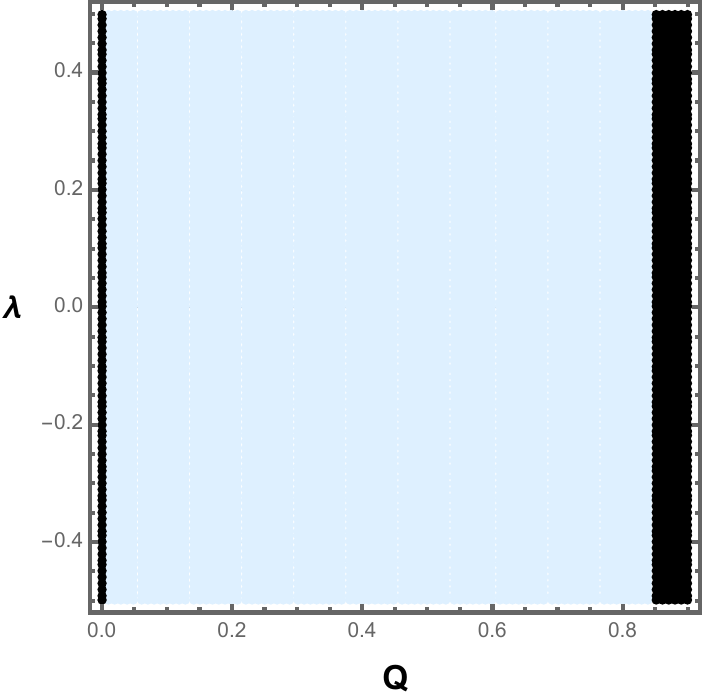}%
            \label{subfig:a}%
        }
        \hspace{1em}
        \subfloat[$R_{\mu\nu}^{2}$]{%
            \includegraphics[width=.4\linewidth]{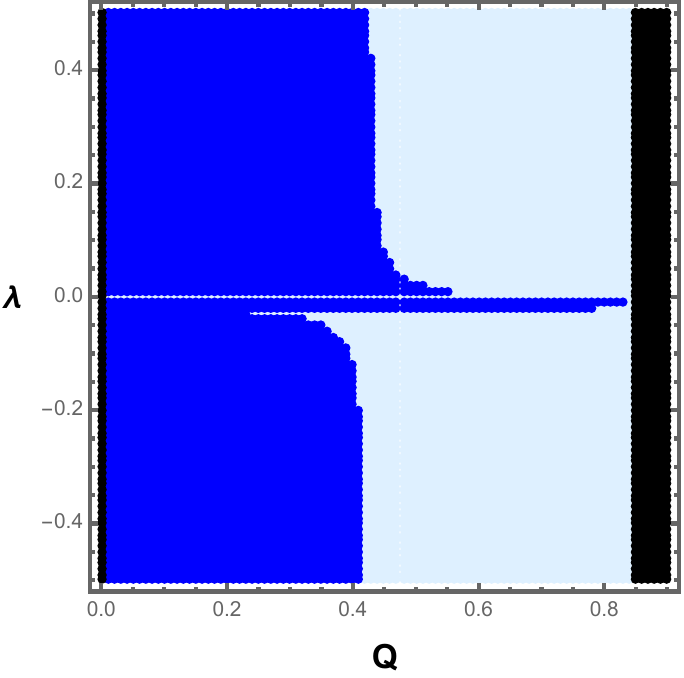}%
            \label{subfig:b}%
        }\\
        \subfloat[$R_{\mu\nu\rho\sigma}^{2}$]{%
            \includegraphics[width=.4\linewidth]{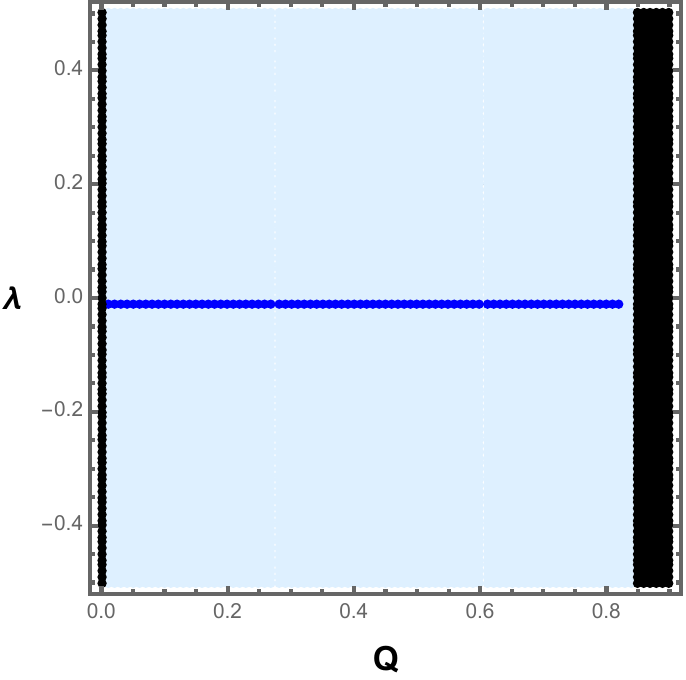}%
            \label{subfig:c}%
        }
        \hspace{1em}
        \subfloat[$C_{\mu\nu\rho\sigma}^{2}$]{%
            \includegraphics[width=.4\linewidth]{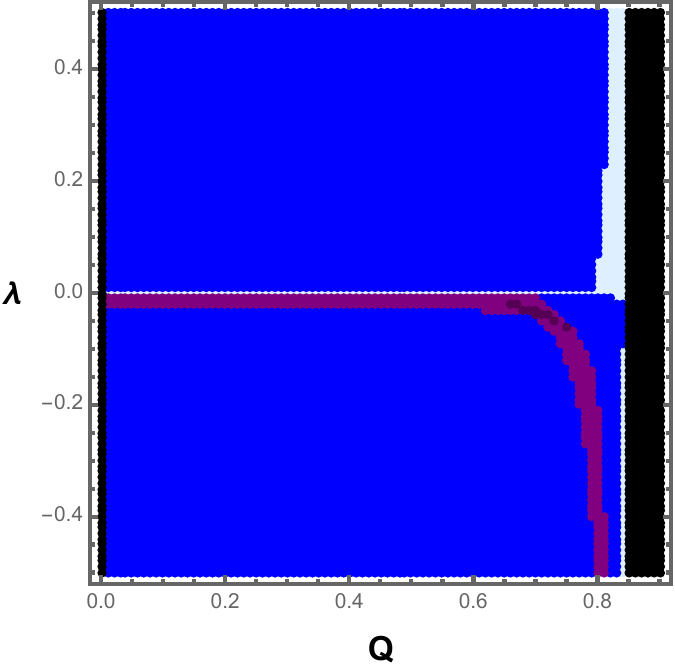}%
            \label{subfig:d}%
        }
        \caption{Parameter scan to determine the number of maxima in the effective potential for different curvature invariants in $F_{1}=1+\lambda$(scalar). Light blue regions denote when $U$ has a single maxima, dark blue regions represent potentials with two local maxima, purple regions represent potentials with three maxima, and dark purple regions show four maxima. Notice the $R^{2}$ curvature invariant is trivial, having only a single maximum (as is the case for complexity=volume). Meanwhile, $C_{\mu\nu\rho\sigma}^{2}$ produces potentials which can have up to four maxima. The right-most black regions represent disallowed values of charge, when the black hole becomes extremal and $U<0$. (Here we set $L=1$ and $M=1$).}
        \label{fig:parameter scans}
    \end{figure}

We analyze the potential (\ref{eq:extremPot}) numerically for a range of $\lambda$ and $Q$, see Figure \ref{fig:parameter scans}. Generally, we find the potential $U$ will have at least one maxima for all parameter values when $r_{+}\neq r_{-}$ (away from extremality). This fact was also observed in \cite{Wang:2023eep}.
 It is easy to see that we will always have at least one maxima between the two horizons since the blackening factor becomes zero at the horizons ($f(r_+)=f(r_-)=0$) and remains negative between the two. Thus, the potential $U(r)$ is positive and has a maximum between $r_-$ and $r_+$ and is negative everywhere else. Since a single maximum is the only requirement for the linear growth of generalized complexity, we find for \emph{all} parameter values there is an allowed codimension-1 observable for complexity. Note that always having at least one maxima is a consequence of having multiple horizons. The neutral eternal AdS black hole, for example, has a far more restricted parameter space of allowed observables \cite{Belin:2021bga}.

For non-extremal black holes, we find that the potential can have more than one maxima depending on the scalar functional being used. Figure \ref{fig:parameter scans}
plots the number of maxima of the potential for each scalar over a range of the parameter values $\lambda$ and $Q$. Via the particle in an effective potential $U$ analogy, a given momentum corresponds to some boundary time and how close to the origin the particle can probe ($r_{\text{min}}$). The Cauchy slice on which the generalized complexity is defined extends from $r_{\textup{min}}$ to the boundary $r\rightarrow\infty$. When the boundary time is large $\tau\rightarrow\infty$, the $r_{\textup{min}}$ lies at the maxima of the potential. Thus, the maxima of the potential represents the closest distance to the singularity that the complexity observable can probe. Below we will discuss the physical interpretation of having multiple maximas.

 In the extremal limit, where $r_{+}=r_{-}=r_{E}$ and thus $f(r)=(r-r_{E})^{2}$, the potential (\ref{eq:extremPot}) is non-positive, having no local (positive) maxima. Consequently, for fixed mass $M$, the numerical scan of the potential breaks down at extremal (and superextremal) values of the charge $Q$. In Figure  \ref{fig:parameter scans} this appears as the maximum charge (for a given mass).\footnote{The extremal horizon radius $r_{E}$ and charge $Q_{E}$ can be found from $f(r_E)=f' (r_E)=0$, i.e., when the blackening factor has a double root. For $L=1$ one has
$$r_E = \frac{\sqrt[3]{6} \left(\sqrt{81 M^2+6}+9 M\right)^{2/3}-6^{2/3}}{6 \sqrt[3]{\sqrt{81 M^2+6}+9 M}}$$
and $Q_E= \sqrt{M r_E + r_E^{4}}$. For $M=1$, as in Figure \ref{fig:parameter scans}, we have $Q_{E}\approx 0.84$.}

\begin{figure}[t!]
    \centering
    \subfloat[innermost maxima probed]{%
            \includegraphics[width=.4\linewidth]{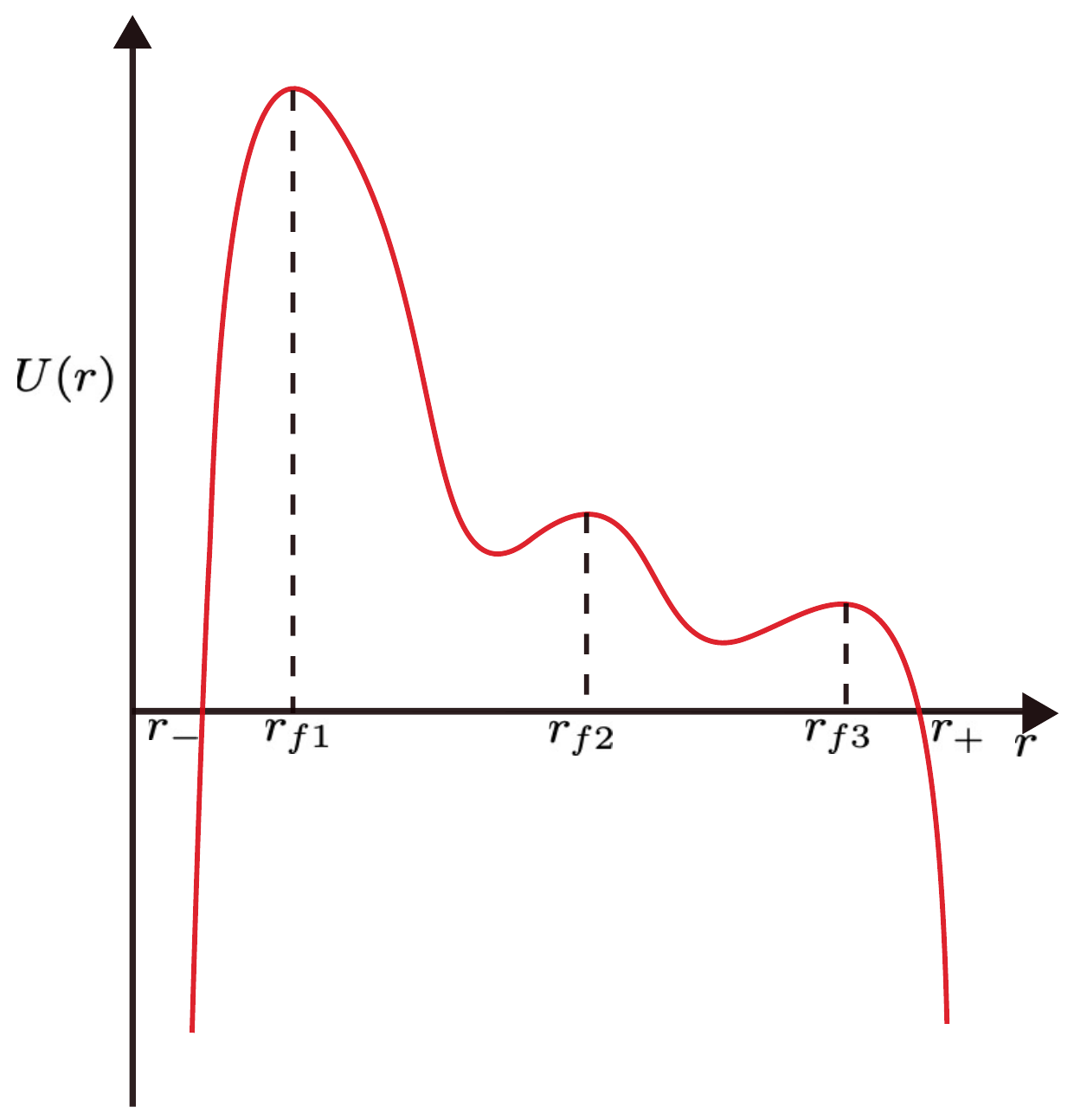}%
            \label{subfig:aa}%
        }
        \subfloat[innermost maxima not probed]{%
            \includegraphics[width=.4\linewidth]{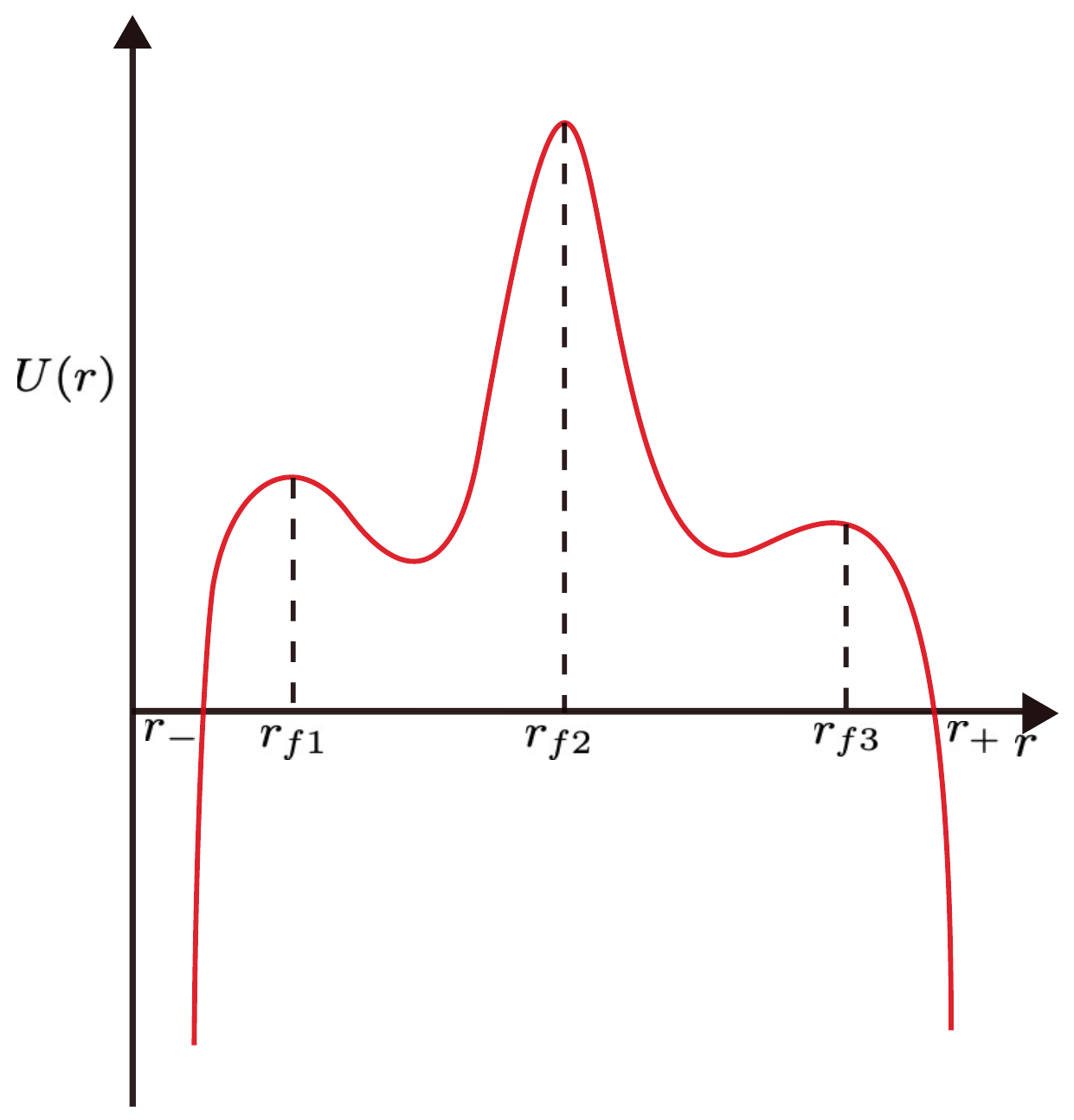}%
            \label{subfig:bb}%
        }
    \caption{\small Effective potential with multiple maxima. The potentials we generate from our calculations have maxima that differ by many orders of magnitude, hence we have used a hand-drawn figure for clarity which still preserves all the features we need. \textbf{Left:} All local maxima are probed by boundary-anchored surfaces. \textbf{Right:} the innermost maximum cannot be probed due to the presence of a larger maximum, which acts as a gravitational barrier.}
    \label{fig:potential cartoon}
\end{figure}

For a given functional to be a candidate for complexity, however, the functional must display linear growth at \emph{late} times. To ensure this is the case, one must look at the surfaces at late boundary times, $\tau\rightarrow\infty,$ and 
the corresponding height of the maxima gives the growth of the complexity observable at late times. For this reason, it only makes sense to count the maxima in descending order as $r$ increases. Thus, the Cauchy surface coming in from infinity can probe larger maxima behind the first one but not see a smaller maxima hidden behind a bigger one. Our plots count the maxima after taking this ordering into account. Figure \ref{fig:potential cartoon} shows an example of the potential.

\subsection{Interpretation of subleading local maxima} \label{sec:interpofsub}

Let us now provide possible interpretations of the multiple maxima of the effective potential $U(r)$ (\ref{eq:extremPot}) characterizing the family of codimension-1 observables.

A common interpretation of complexity=anything is that specifying the function $a(r)$ helps fix the ambiguities in the boundary definition of complexity, including the choice of elementary gates and cost functions that define the dual quantum circuit. Each surface corresponds to a turning point $r = r_{\textup{min}}$ where $\dot{r} = 0$ and thus $P_v^2 = U(r_{\textup{min}})$. Solving $P_v^2 = U(r_{\textup{min}})$ determines the allowed turning radii for any boundary time. To find the allowed turning radii $r_f$ for boundary time $\tau\rightarrow\infty$ we solve for the maxima of the potential. Thus, multiple local maxima of $U(r)$ yield multiple solutions for $r_{f}$ and, therefore, several candidate extremal slices, provided there are no higher barriers in the potential separating the local maximum in question from the AdS boundary. Figure~\ref{fig:potential cartoon} gives illustrative examples. On the left, we show a scenario with three local maxima, each defining a different extremal slice. On the right, we show three local maxima again; however, one of them is hidden behind a larger maximum and therefore cannot be probed by boundary-anchored slices.
Regardless of specific features in the potential, the existence of several maxima opens the question of how we interpret them from a boundary perspective. We elaborate on this issue below.

\bigskip

\noindent\textbf{Interpretation 1: Distinct locally optimal circuits for the full state.} In our first interpretation, \emph{each} local maximum can be viewed as a distinct, locally optimal way to prepare \textit{the same} final state, but with different gate arrangements and thus different computational complexity $\mathcal{C}_i(t)$. One of these maxima corresponds to the \emph{optimal} or most efficient circuit, the one that \emph{minimizes} the number of gates in the circuit\footnote{Maximizing the (generalized) volume corresponds to minimizing the number of gates needed to prepare the quantum state (based on their cost), i.e., the most efficient circuit. This follows from the Lorentzian min-flow/max-cut theorem for Lorentzian manifolds \cite{Headrick:2017ucz}, along with the `Lorentzian threads' reinterpretation of Complexity=Volume \cite{Pedraza:2021mkh,Pedraza:2021fgp}. For Complexity=Anything, a generalized min-flow/max-cut theorem still holds but is formulated more rigorously in terms of measure theory rather than vector flows \cite{Caceres:2023ziv}.}
\[
\mathcal{C}(t) \;=\; \min \{ \mathcal{C}_1(t), \mathcal{C}_2(t), \ldots\}\,.
\]
Since $\mathcal{C}_i(t)$ is obtained by \emph{maximizing} the generalized volume, this construction yields a `minimax' prescription, and it is completely analogous to the `maximin' prescription for (covariant) entanglement entropy \cite{Wall:2012uf}.
Although other maxima yield a stationary (extremal) generalized volume, they correspond to suboptimal paths that reach the \emph{same} final state with a larger gate count. From a quantum-circuit perspective, this interpretation is natural, as there may be various ``locally optimal'' routes in gate space, each achieving the same state but at different circuit depths and different total gate counts. The smallest extremal value gives the \emph{dominant} or \emph{most efficient} contribution to complexity, while other maxima represent alternative preparations that are more expensive, yet still locally extremal.

\medskip

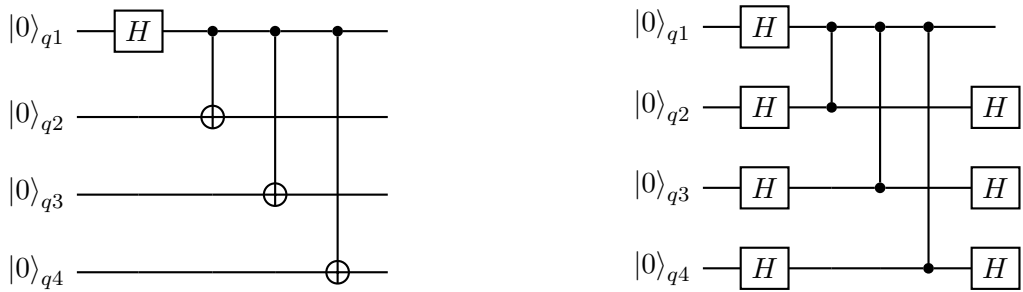
\begin{figure}[t!]
\centering
\begin{minipage}[t]{0.45\textwidth}
\centering
\begin{quantikz}[row sep=0.7cm, column sep=0.5cm]
\lstick{$\ket{0}_{q1}$} & \gate{H} & \ctrl{1} & \ctrl{2} & \ctrl{3} & \qw \\
\lstick{$\ket{0}_{q2}$} & \qw      & \targ{}  & \qw      & \qw      & \qw \\
\lstick{$\ket{0}_{q3}$} & \qw      & \qw      & \targ{}  & \qw      & \qw \\
\lstick{$\ket{0}_{q4}$} & \qw      & \qw      & \qw      & \targ{}  & \qw
\end{quantikz}
\hspace{5cm}
\end{minipage}
\hfill
\begin{minipage}[t]{0.45\textwidth}
\centering
\begin{quantikz}[row sep=0.52cm, column sep=0.5cm]
\lstick{$\ket{0}_{q1}$} & \gate{H} & \ctrl{1} & \ctrl{2} & \ctrl{3} & \qw      \\
\lstick{$\ket{0}_{q2}$} & \gate{H} & \control{} & \qw      & \qw      & \gate{H} \\
\lstick{$\ket{0}_{q3}$} & \gate{H} & \qw      & \control{} & \qw      & \gate{H} \\
\lstick{$\ket{0}_{q4}$} & \gate{H} & \qw      & \qw      & \control{} & \gate{H}
\end{quantikz}
\end{minipage}
\caption{\small Side-by-side comparison of two locally optimal circuits generating the 4-qubit GHZ state 
$\tfrac{1}{\sqrt{2}}(|0000\rangle + |1111\rangle)$. 
\textbf{Left:} A CNOT-based approach. 
\textbf{Right:} A CZ-based approach, common on certain hardware where CZ is the native two-qubit gate. Each uses three two-qubit gates (CNOT or CZ), the minimum needed to entangle four qubits in GHZ form.}
\label{fig:GHZ_comparison}
\end{figure}

\noindent\emph{Example:} Imagine we want to prepare the 4-qubit GHZ state from an unentangled state, with Hadamard, CNOT, and Controlled-Z gates, e.g., 
\[
\ket{0000}\rightarrow \frac{1}{\sqrt{2}}(|0000\rangle + |1111\rangle)\,.
\]
There are several ways to construct this state, and ultimately the optimal circuit depends on both the total number of gates and the cost of executing each type of gate. In Figure \ref{fig:GHZ_comparison} we show a side-by-side comparison of two such circuits. In the first circuit, the CNOT gate is assumed to have a low implementation cost, making it the natural choice; therefore, we can prepare the 4-qubit GHZ with just one Hadamard and three CNOTs (4 total gates). This is minimal for that specific gate set. In the second circuit, by contrast, the CNOT gate is costly, so we use single-qubit rotations and controlled-Z (CZ) gates. To emulate a CNOT via CZ, we insert extra Hadamards on the target qubit, thus increasing the total gate count (10 total gates). Although the second approach uses more single-qubit gates, these are cheap in many hardware platforms and can often be executed in parallel. From a ``hardware scheduling'' viewpoint, the time cost is still dominated by the three CZs; the extra Hadamards can be inserted before or after each CZ with almost no additional depth. Ultimately, the optimal circuit will depend on the specific cost associated with each gate type and the overall gate arrangement.

Although subdominant solutions do not change the \emph{leading} value of complexity and its growth rate, they can become relevant if boundary conditions or external constraints cause the globally dominant solution to lose its primacy. For instance, a ``phase transition'' in which a secondary branch overtakes the original most efficient solution is possible if the cost functionals associated with different gates vary over time.

\medskip

\noindent
\textbf{Interpretation 2: Additive contributions from independent subsectors.}
An alternative interpretation is the following: \emph{all} local maxima contribute \emph{additively} to the total complexity, much like summing distinct saddle-point contributions in a path integral. One can imagine each extremal slice as independent channels through which gates accumulate:
\[
\mathcal{C}(t) \;=\; \sum_i \alpha_i\, \mathcal{C}_i(t),
\]
where $\mathcal{C}_i(t)$ is the contribution from the $i^\text{th}$ local maximum for generalized volume and $\alpha_i$ are weighing constants that may arise from the measure of the path integral.\footnote{Within this interpretation, one may even account for contributions from maxima in the potential that are concealed behind a barrier (see Fig. \ref{fig:potential cartoon}b). Although these correspond to surfaces not anchored at the AdS boundary, they could be understood as `complexity islands,' disconnected contributions that, much like the so-called `entanglement islands,' still play a role in the calculation of the associated observable.} From a quantum circuit perspective, this suggests that if the full system splits into multiple blocks or subsectors that evolve nearly independently (so that no strong cross-gates are needed), the total circuit cost is simply the sum of each block's complexity. In that sense, subleading extremal slices do not remain subdominant: they represent distinct sectors whose complexities add together.  Even in a single connected system, one can still conceive of partial or quasi-isolated channels of evolution that collectively determine the total cost.

\medskip

\begin{figure}[t]
\centering
\begin{quantikz}[row sep=0.75cm, column sep=0.75cm]
\lstick{{\color{blue}$\ket{0}_{q1}$}}
& \gate{H}
& \ctrl{1}
& \ctrl{2}
& \qw
\\
\lstick{{\color{blue}$\ket{0}_{q2}$}}
& \qw
& \targ{}
& \qw
& \qw
\\
\lstick{{\color{blue}$\ket{0}_{q3}$}}
& \qw
& \qw
& \targ{}
& \qw
\\[1em]
\lstick{\color{red}$\ket{0}_{q4}$}
& \gate{H} & \ctrl{1}   & \qw        & \qw \\
\lstick{\color{red}$\ket{0}_{q5}$}
& \gate{H} & \control{} & \ctrl{1}   & \qw \\
\lstick{\color{red}$\ket{0}_{q6}$}
& \gate{H} & \qw        & \control{} & \qw
\end{quantikz}
\caption{\small
Two independent 3-qubit subcircuits (blue and red).
\textbf{Top}: qubits $(q_1,q_2,q_3)$ form a standard GHZ circuit: 
Hadamard on $q_1$, followed by CNOTs ($q_1 \!\to\! q_2$ and $q_1 \!\to\! q_3$). 
\textbf{Bottom}: qubits $(q_4,q_5,q_6)$ each receive a Hadamard, then two CZ gates 
between $(q_4,q_5)$ and $(q_5,q_6)$ introduce phases in the amplitudes where both qubits are 1, 
yielding a multi-qubit superposition in this sector.}
\label{fig:twosubsectors}
\end{figure}
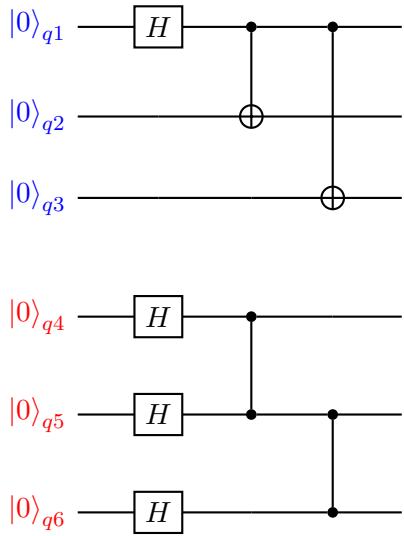

\noindent\emph{Example:} Consider a 6-qubit system divided into two independent 3-qubit subsectors, labeled A and B. For definiteness, let subsector A comprise qubits $\{1,2,3\}$ and subsector B comprise qubits $\{4,5,6\}$. We impose a conserved quantum charge (e.g.\ a $U(1)$ charge or a topological invariant) that forbids any operations coupling A and B within the chosen elementary gate set. In other words, each subsector carries a distinct charge quantum number, and the available gates preserve that charge. Any entangling gate acting on one qubit from A and one from B would exchange charge between sectors and thus violate the symmetry, so cross-sector entangling gates are disallowed. This superselection rule ensures the total evolution factorizes as $U_{\text{total}} = U_A \otimes U_B$, where $U_A$ acts only on sector A and $U_B$ acts only on sector B. In Figure \ref{fig:twosubsectors} we show a concrete example of such a circuit, with Hadamard, CNOT, and CZ gates. The first subsector is prepared as
\[
\ket{000}_A \;\longrightarrow\; \frac{1}{\sqrt{2}}\bigl(\ket{000} + \ket{111}\bigr),
\]
while the second follows
\[
\ket{000}_B \;\longrightarrow\; \frac{1}{2^{3/2}}
\Bigl(\,
|000\rangle 
+ |001\rangle 
+ |010\rangle 
- |011\rangle
+ |100\rangle 
+ |101\rangle 
- |110\rangle 
+ |111\rangle
\Bigr).
\]
We assume CNOT gates are cost-efficient in the first subsector, while CZ gates are cost-efficient in the second. Because the chosen gates and costs leave the two subsectors effectively decoupled, the total complexity is simply the sum of their individual complexities.

\bigskip

\noindent
\textbf{Connections to previous work.} Various authors have noted that, while multiple extremal surfaces generally exist in the complexity=anything framework, they are often discarded as subleading \cite{Wang:2023eep,Jiang:2023jti,Wang:2023noo}.\footnote{We note, however, that previous proposals adopt a `maximax' prescription instead of the 'minimax' proposed here, which, given the quantum-circuit interpretation above, we find more physically appealing.} Nevertheless, the idea that subdominant surfaces can carry physical significance is well-established in holography. One prominent example is \emph{entwinement} \cite{Balasubramanian:2014sra}, which encodes the entanglement of internal degrees of freedom not captured by the minimal entangling Ryu–Takayanagi (RT) surface \cite{Balasubramanian:2016xho,Balasubramanian:2018ajb,Lin:2016fqk,Craps:2022pke}. In spherical AdS black holes, for instance, the non-minimal extremal surfaces have areas that are interpreted as entwinement and can probe the ``entanglement shadow'', a region inaccessible to the global minimal surface \cite{Freivogel:2014lja}. By analogy, our \emph{Interpretation 1} suggests that additional locally optimal bulk slices ---though giving rise to a larger complexity--- may represent legitimate alternative circuits for preparing the \emph{same} final state, just as subleading RT surfaces can encode physically relevant information about internal degrees of freedom. Hence, these slices are ``subdominant'' only in a global sense but remain meaningful for exploring certain aspects of the state’s structure.

On the other hand, our \emph{Interpretation 2} parallels how subleading geodesics can contribute nontrivially to correlation functions of heavy operators. In the geodesic approximation to two-point functions, one must in principle sum over \emph{all} geodesic saddles—both the minimal path and any longer winding geodesics \cite{Balasubramanian:1999zv,Louko:2000tp}. In AdS$_3$ with conical defects, for example, these subleading geodesics provide essential corrections that restore the correct correlator behavior \cite{Arefeva:2014aoe,Arefeva:2016wek,Arefeva:2016nic,Berenstein:2022ico}. Likewise, if the system is effectively divided into quasi-isolated subsectors (with no entangling gates across them), each sector can contribute independently to the total complexity, mirroring the sum over multiple geodesic saddles. Similar scenarios arise in multi-boundary wormholes, where multiple saddles (or extremal slices) appear \cite{Balasubramanian:2014hda}. Indeed, certain proposals such as ``binding complexity'' explicitly decompose total complexity into additive pieces associated with separate subsystems \cite{Balasubramanian:2018hsu}. In a similar vein, one might imagine a refinement of holographic complexity that includes secondary $1/N$ corrections, e.g., a sum over topologies reminiscent of a path integral. Such proposals have recently been explored to investigate the late-time plateau of complexity \cite{Iliesiu:2021ari,Miyaji:2025yvm}.

\bigskip

\noindent
\textbf{Summary.}
In summary, the existence of additional local maxima in the effective potential admits two natural interpretations:
\begin{enumerate}
\item \emph{Locally optimal yet subdominant circuits for the same final state.} These surfaces satisfy the extremality condition but represent a less efficient circuit than the optimal one. While they do not modify the leading (dominant) complexity, they may become relevant under certain conditions. Such suboptimal complexities can still encode extra information about the state, mirroring the role of entwinement in holography.
\item \emph{Independent channels in an additive sum.} If the quantum system is truly factorized or effectively decoupled under the chosen gates and cost functions, each local maximum can be regarded as a separate contribution to the total complexity, reminiscent of summing multiple geodesics when computing a two-point function of heavy operators.
\end{enumerate}
Both perspectives enrich the complexity = anything framework: the first highlights hidden substructures tied to particular paths in the complexity geometry, while the second raises the intriguing possibility that holographic complexity, like other bulk observables, may involve sums over multiple extremal configurations. Ultimately, determining which interpretation applies (or whether both are relevant) depends on the detailed bulk-to-boundary map for circuit complexity. Establishing that map ---perhaps by including quantum corrections, by ensemble averaging over topologies, or by performing a first-principles derivation--- is an important direction for future research.

\section{Complexity=anything in 2D gravity} \label{sec:CA2D}

In this section we develop the complexity=anything proposal for two-dimensional theories of dilaton gravity. To build intuition, we first dimensionally reduce the codimension-1 observables for generalized complexity in four-dimensional Einstein gravity. We then find a more general class of codimension-1 observables for complexity in two-dimensions using an analog of the extremization problem described in the introduction.

\subsection{Dimensional reduction of complexity=anything}

Assume the four-dimensional metric is spherically symmetric. As an ansatz, let the line element have the form
\begin{equation}\label{eq:metricansatz}
    ds^2=g_{\mu\nu}dx^\mu dx^\nu +\ell^{2}\Phi(x) d\Omega_{2}^2, 
\end{equation}
for two-dimensional metric $g_{\mu\nu}$ with $\mu,\nu=0,1$, $\Phi(x)$ is the (dimensionless) dilaton, and $\ell$ is some four-dimensional length scale. Here we assume $\Phi=\Phi(r)$. Upon substitution into the Einstein-Hilbert (-Maxwell) action and integrating over the 2-sphere one finds (see, e.g., the appendices of \cite{Svesko:2022txo,Galante:2025tnt})\footnote{ The boundary action is the usual Gibbons-Hawking-York (GHY) term. This arises via spherical dimensional reduction of the GHY term for Einstein gravity, where the induced metric at the boundary obeys Dirichlet boundary conditions. Our analysis applies equally well for theories which obey different boundary conditions, for which the reduced one-dimensional boundary action differs, cf. \cite{Galante:2025tnt,Banihashemi:2025qqi}.}
\beq I_{2D}=\frac{4\pi \ell^{2}}{16\pi G_{4}}\int_{\mathcal{M}}d^{2}x\sqrt{-g}\left[\Phi R+V(\Phi)\right]+\frac{4\pi \ell^{2}}{8\pi G_{4}}\int_{\partial\mathcal{M}}dy\sqrt{-h}\Phi K\;.\label{eq:2Dactgen}\eeq
Here, $R$ denotes the Ricci scalar associated with the two-dimensional metric $g_{\mu\nu}$, and $K$ is the trace of the extrinsic curvature associated with the one-dimensional induced metric $h_{\mu\nu}$. We leave the dilaton potential $V(\Phi)$ generic. Its precise form depends on the higher-dimensional solution from whence it came and each specific form corresponds to specific theories, e.g.,  AdS$_{2}$-JT gravity has $V(\Phi)=2\Phi/L_{2}^{2}$ for AdS$_{2}$ length $L_{2}$.

Similarly, let us perform a dimensional reduction of the codimension-1 observables \eqref{eq:cgencod1}. Substituting in the ansatz \eqref{eq:metricansatz} and integrating over the sphere gives\footnote{Here it is useful to know $\int_{\Sigma}d^{3}y\sqrt{h_{3D}}=4\pi \ell^{2}\int_{\Sigma}d y\sqrt{h_{1D}}\Phi$. The factor of $\Phi^{3/4}$ is because one additionally performs a Weyl transformation $g_{\mu\nu}\to \omega^{2} g_{\mu\nu}$ for $\omega=\Phi^{-1/4}$ to remove unwanted kinetic terms for the dilaton that arise from the dimensional reduction of four-dimensional gravity. A similar transformation is needed in higher than four dimensions.}
\begin{equation}
    \mathcal{C}_{\text{gen}}=\max\frac{1}{G_{2} L}\int_{\Sigma}d\sigma\sqrt{h}\Phi(r)^{3/4}F_{1}
\label{eq:Cgen2Ddim}\end{equation}
where  
$\sigma$ is the single radial coordinate on the slice $\Sigma$, $G_{2}^{-1}=4\pi\ell^{2}/G_{4}$ is the effective two-dimensional Newton's constant, and $F_{1}=a(r)$ is the dimensionally reduced functional.
When $F_{1}=1$, the complexity functional (\ref{eq:Cgen2Ddim}) describes the complexity=volume proposal explored in \cite{Carrasco:2023fcj} (see also \cite{Schneiderbauer:2019anh,Bhattacharya:2023drv,Aguilar-Gutierrez:2023tic,Fu:2024vin}), when approximating $\Phi\approx \phi_{0}+\phi$ for $\phi_{0}\gg \phi$, as is the case for JT gravity.\footnote{In \cite{Brown:2018bms}, volume complexity took the same form as (\ref{eq:Cgen2Ddim}) except $\Phi$ was treated as a large constant.} As in the higher-dimensional case, the generalized complexity functional (\ref{eq:Cgen2Ddim}) describes deviations away from the analog of volume complexity in two-dimensions, i.e., $F_{1}=1+\lambda(\text{scalar})$, where now ``scalar'' refers to the dimensional reduction of the quadratic curvature invariants considered in higher dimensions (we report the dimensionally reduced curvature invariants in Appendix \ref{app:dimredx}).
In particular, we find $F_{1}$ is a specific functional of the dilaton $\Phi$, covariant derivatives of $\Phi$, possibly non-minimally coupled to two-dimensional curvature invariants, i.e., powers of the two-dimensional Ricci scalar.

Following the spirit of \cite{Belin:2021bga}, we can write down an effective one-dimensional classical mechanics problem with generic Lagrangian $\mathcal{L}_{\text{gen}}=\sqrt{h}\Phi^{3/4}a(r)$. This would lead to a particle in an effective potential $U(r)$ that now depends on the dilaton. As before, linear growth in the complexity requires the potential $U$ have at least one local maxima. Rather than pursuing this top-down approach -- which we are guaranteed to recover the same physics as in four-dimensions described in Section \ref{sec:CA4D}, just now cast in two-dimensional variables -- let us instead develop a two-dimensional complexity=anything proposal from the bottom-up.

\subsection{Two-dimensional complexity=anything: bottom-up}

Let us now take a `bottom-up' approach to construct a family of codimension-1 observables (expected to be) dual to complexity for a general class of dilaton theories characterized by the action (\ref{eq:2Dactgen}) with generic dilaton potential $V(\Phi)$. No restrictions will be imposed on the form of the potential. We comment on the physical implications of this in Section \ref{sec:disc}.

The metric and dilaton equations of motion are, respectively,
\begin{equation}\label{eq:dilaton_theory_eom}
    \begin{split}
        \nabla_\mu \nabla_\nu\Phi -g_{\mu\nu}\Box\Phi+\frac{1}{2} g_{\mu\nu}V(\Phi)=0&,\\
        R=-\partial_\Phi V(\Phi)&\;.
    \end{split}
\end{equation}
A general solution to the equations of motion comes from using a covariant B{\"a}cklund transformation (see, e.g., \cite{Cavaglia:1998xj})\footnote{Here we follow the conventions of \cite{Aguilar-Gutierrez:2024nst} in Lorentzian signature.}
\begin{equation}\label{eq: 2dim metric}
    \dd s^2=-N(r)\dd t^2+\frac{\dd r^2}{N(r)},\quad \Phi=\Phi_p \frac{r}{L},
\end{equation}
where $\Phi_p$ is a positive dimensionless constant. Since $R=-\partial^{2}_{r}N$, the dilaton equation of motion gives
\begin{equation}
    N(r)=\frac{L}{\Phi_r}\int_{r_h}^r d\bar{r} V(\bar{r})\;.
\end{equation}
The metric (\ref{eq: 2dim metric}) describes a two-dimensional Lorentzian geometry where $N(r_{h})=0$, such that $r=r_{h}$ characterizes a 2D black hole horizon. Here we will be primarily interested in the case of AdS$_{2}$-JT gravity, 
where the geometry is the AdS$_{2}$ `black hole', $N(r)=(r^{2}-r_{h}^{2})/L_{2}^{2}$.\footnote{The AdS$_{2}$ black hole is really AdS$_{2}$-Rindler space and thus does not have a curvature singularity itself. In 2D dilaton-gravity models, the curvature singularity of the higher-dimensional black hole corresponds to $\Phi\to-\infty$, i.e., in regions of strong gravity.}

We now follow the spirit of \cite{Belin:2021bga} and seek general candidates for holographic complexity, and show that such proposals feature late time linear growth. Specifically, 
we propose the following family of codimension-1 observables
\begin{equation}
    \mathcal{C}_{\textup{gen}}=\frac{1}{G_2 L}\int_{\Sigma_{F_2}} dy \sqrt{h}F_1(g_{\mu\nu},\Phi, X^{\mu}),
\label{eq:complexfam2d}\end{equation}
with $F_1$ and $F_{2}$ arbitrary functions of the metric and dilaton, $h$ is the determinant of the induced metric on the codimension-1 surface $\Sigma_{F_2}$, and $X^{\mu}(y)$ describes the embedding of the codimension-1 surface. Generally, $F_{1}\neq F_{2}$, and $\Sigma_{F_{2}}$ is fixed asymptotically such that $\partial\Sigma_{F_{2}}=\sigma_{\text{CFT}}$, a boundary timeslice (we imagine working in an asymptotically AdS$_{2}$ background, or, more generally, a bulk 2D spacetime with a timelike boundary). The bulk slice $\Sigma_{F_{2}}$ can be found in a diffeomorphism invariant way via the extremization prescription
\begin{equation}
    \delta_{X}\left(\int_{\Sigma} dy \sqrt{h}F_2(g_{\mu\nu},\Phi, X^{\mu})\right)=0\;.
\end{equation}
Observe that for $F_{1}=F_{2}=\Phi$ (or a positive power thereof), one recovers the complexity=volume proposal for JT gravity \cite{Carrasco:2023fcj} (see also \cite{Schneiderbauer:2019anh,Bhattacharya:2023drv,Aguilar-Gutierrez:2023tic,Fu:2024vin}).

Let us now check that our proposal for generalized complexity (\ref{eq:complexfam2d}) obeys late time linear growth. Moving forward, we assume $F_{1}=F_{2}$.  
An arbitrary hypersurface in the two dimensional background \eqref{eq: 2dim metric} can be parametrized by a single coordinate variable $y$
with induced metric
\begin{equation}
    h=g_{\mu\nu}\partial_y x^\mu \partial_y x^\nu=-t'(y    )^2N(r)+\frac{r'(y)^2}{N(r)}\;.
\end{equation}
with $t'(y),\ r'(y)$ denoting the derivative with respect to $y$ of the surface coordinates $t$ and $r$, respectively. The observable (\ref{eq:complexfam2d}) then reads
\begin{equation}\label{eq: observable F1=F2}
    \mathcal{C}_{\text{gen}}=\frac{1}{G_2 L}\int_{\Sigma} dy \sqrt{-(t')^2N(r)+\frac{(r')^2}{N(r)}}a(r)\equiv\frac{1}{G_{2}L}\int_{\Sigma}dy\mathcal{L}\;,
\end{equation}
where $a(r)$ is $F(\Phi,g_{\mu\nu})$ evaluated over the surface $\Sigma$. 
As in higher-dimensions finding the extrema of this integral function reduces to a one-dimensional Euler-Lagrange problem with Lagrangian $\mathcal{L}$.

Further simplifying this, we remark, that since our problem is reparametrization invariant, it is possible to set a concrete gauge. In particular, we take
\begin{equation}\label{eq: constraint}
    \sqrt{-(t')^2 N(r)+\frac{(r')^2}{N(r)}}=a(r)\;.
\end{equation}
Studying the Euler-Lagrange equations, since the effective Lagrangian $\mathcal{L}$ is independent of $t$, there exists a conserved momentum $P_{t}$
\begin{equation}
    P_t\equiv \frac{\partial \mathcal{L}}{\partial t'}=-\frac{t' N(r) a(r)}{\sqrt{-(t')^2 N(r)+\frac{(r')^2}{N(r)}}}=-t' N(r)\;,
\end{equation}
where in the second equality we used the gauge fixing (\ref{eq: constraint}). We immediately find 
\begin{equation}\label{eq: eq for t}
    t'=-\frac{P_t}{N(r)}\;,
\end{equation}
from which, via the constraint \eqref{eq: constraint}, we find the equation of motion for $r'$ is also fixed:
\begin{equation}
(r')^2-a(r)^2N(r)=P_t^2\;. 
\end{equation}
As in higher dimensions, we may interpret this differential equation as characterizing a particle in a potential $U\equiv-a(r)^2N(r)$. 
In this way, the $r$ coordinate of the line runs from a minimal value $r_{\text{min}}$, located at the point where $U(r_{\textup{min}})=P_t^2$ (where the ``kinetic energy'' vanishes), up to infinity. Note that $r_{\textup{min}}$ depends on the value of $P_t$. 

To find the complete profile of the surface, we substitute the solution $r(y)$ into \eqref{eq: eq for t} and taking into account that, by symmetry, the minimum value of the coordinate $r$ lies at the turning point located at the time slice $t=0$ (not to confuse with the time at which the surface hits the boundary, which is denoted by $\tau$). The time at which the surface crosses the boundary of AdS$_2$ is 
\begin{equation}\label{eq: time}
    \tau=(t_R+t_L)=2t_R=2\int dt= 2\int dr\frac{dy}{dr}\frac{dt}{dy}=-\int_{r_\textup{min}}^\infty dr \frac{2P_t}{N(r)\sqrt{P_t^2-U(r)}}.
\end{equation}
Let us now show that, if the potential $U(r)$ has a (local) maximum inside the horizon, the observable will exhibit late linear growth. If there exists a single maximum located at $r_f$, then
\begin{equation}
    U(r_f)\equiv P_\infty^2,\quad U'(r_f)=0,\quad U''(r_f)\leq 0\;.
\end{equation}
Here $P_\infty$ is a constant that coincides with the value of $P_t$ of the surface that ends at $\tau\rightarrow \infty$. Indeed, the denominator in the integral \eqref{eq: time} can be expanded up to second order near the maximum setting $P_t=P_\infty$
\begin{equation}
    P_\infty^2-U(r)\approx -\frac{1}{2} U''(r_f)(r-r_f)^2.
\label{eq:Pinfty2D}\end{equation}
Substituting this back into \eqref{eq: time}, the integral has a non-integrable singularity and $\tau$  diverges to positive infinity. 
Thus, we have a family of surfaces parametrized by $P_t$ defined for all times between $t\in [0,\infty)$. 

All that remains is showing the observable \eqref{eq: observable F1=F2}  defined on these surfaces grows linearly  for large $\tau$. As in \cite{Belin:2021bga}, we have
\begin{equation}
    \frac{d \mathcal{C}_{\textup{gen}}}{d \tau}= \left.\frac{P_t}{2}\right|_{\partial \Sigma}
\end{equation}
On the other hand, the (inverse) of the time $\tau$ derivative of $P_{t}$ is
\begin{equation}
    \frac{d \tau}{d P_t} = \int_{r_{\textup{min}}}^{\infty}dr\frac{2U(r)}{N(r)\sqrt{(P_t^2-U(r))^3}}+\frac{d r_{\textup{min}}}{d P_t}\left[\frac{2P_t}{N(r)\sqrt{P_t^2-U(r)}}\right]_{r\rightarrow r_{\textup{min}}}\;.
\end{equation}
 Both terms are divergent near the limiting case $P_t \rightarrow P_\infty$ or, equivalently, $r_{\textup{min}}\rightarrow r_{f}$. The dominant divergence at late times comes from the first term, 
\begin{equation}
    \frac{d \tau}{d P_t} = -\frac{2\sqrt{2}U(r_f)}{N(r_f)\sqrt{-U''(r_f)^3}(r_{\textup{min}}-r_f)^2},
\end{equation}
where we substituted in expansion (\ref{eq:Pinfty2D}) and performed the integral. Since $d\tau/dP_{t}$ diverges as $r_{\text{min}}\to r_{f}$, $P_{t}$ approaches the constant value $P_{\infty}$. As such, our proposal (\ref{eq:complexfam2d}) captures a universal feature of complexity, and thus serves as a candidate for its bulk dual. Further, following the same reasoning in \cite{Belin:2021bga} (reviewed above),  
 our proposal will also exhibit the switchback effect since the one-dimensional observables are additive for two-dimensional shockwave geometries, and the observables evaluated on the final slice has late time linear growth.

We conclude with a final remark. Suppose the potential displays multiple (local) maxima.  
 According to \eqref{eq: time}, different values of $P_t$ (or similarly, different surfaces) might reach the boundary at the same time. For this reason, a degeneracy on the extremal surfaces appears, and an extra condition must be imposed in order to select a concrete surface among all the possibilities (see Section \ref{sec:interpofsub} for possible conditions to impose).

\section{Comments on the holographic dictionary for complexity=anything}\label{sec:dictionary}

The ambiguities in the definition of complexity should be reflected in both bulk and boundary perspectives. While different candidates within the `complexity=anything' proposal can serve as viable bulk observables, this flexibility must also be mirrored in the boundary definition of complexity. In particular, different choices of the bulk observable, denoted here as $W$, should correspond to different measures of complexity in the CFT. 

This connection can be made explicit using the gravitational phase-space formalism. For example, in \cite{Belin:2018bpg}, it was argued that a natural definition of boundary complexity for coherent bulk states (i.e., states that can be prepared using the Euclidean path integral) should be given by a notion of distance in the space of sources. Schematically,
\begin{equation} \label{generalbdydef}
\mathcal{C}(s_i,s_f)=\int_{s_i}^{s_f}ds\, \mathcal{F}[g_{ab}\dot{\lambda}^a\dot{\lambda}^b]\,, 
\end{equation}
where $\lambda^a$ are coordinates on the complexified space of sources, $\dot{\lambda}^a$ are their derivative with respect to an affine parameter in the space of sources, $g_{ab}$ is the (K\"ahler) metric on this space, and $\mathcal{F}[\cdots]$ is a cost function that must be extremized.
Further, to attain the CV proposal, the appropriate cost function was argued to be the kinetic energy, $\mathcal{F}[x]=x$, leading to a `first law' of complexity for small variations of the target state \cite{Belin:2018bpg}
\begin{equation}\label{firstlawCV} \delta_{\lambda_f}\mathcal{C}=\dot{\lambda}^a|_{\lambda_f}g_{ab}\delta \lambda_f^b\,. \end{equation}
Notably,  by reversing the logic, i.e., assuming the CV proposal, this first law (\ref{firstlawCV}) was shown to be equivalent to the linearized Einstein equations around pure AdS \cite{Pedraza:2021mkh,Pedraza:2021fgp,Pedraza:2022dqi}. In essence, requiring the volume of the maximal slice to respond correctly to small changes in the state (in analogy with the first law of entanglement \cite{Lashkari:2013koa,Faulkner:2013ica,Swingle:2014uza,Faulkner:2017tkh,Haehl:2017sot,Agon:2020mvu,Agon:2021tia}) imposes the bulk equations of motion. Hence, gravitational dynamics emerges from `spacetime complexity'.  

Importantly, the arguments of \cite{Pedraza:2021mkh,Pedraza:2021fgp,Pedraza:2022dqi}  extend to more general theories, such as higher-curvature gravities and semi-classical JT gravity \cite{Carrasco:2023fcj}, provided one adopts the appropriate bulk functional.  
Such studies cement the idea that a well-defined boundary complexity (and its variations) can be consistently mapped to a chosen bulk functional $W$. The key to this map is the symplectic structure of the theories.

In the covariant phase-space formalism, the bulk theory is equipped with a symplectic 2-form $\Omega_{\text{bulk}}$ on the space of solutions, and likewise, the boundary CFT has a symplectic form $\Omega_{\text{bdy}}$ encoding its unitary dynamics. A key result in holography is that these symplectic forms coincide for on-shell deformations \cite{Belin:2018fxe}. That is, there is a one-to-one map between bulk deformations and boundary state perturbations such that $\Omega_{\text{bulk}}(\delta_1,\delta_2) = \Omega_{\text{bdy}}(\delta_1,\delta_2)$ for corresponding variations. This fact was used in \cite{Belin:2018bpg} to argue that a concrete notion of boundary complexity can be \emph{defined} via a bulk quantity. In the CV case, for example, the maximal slice volume acts as the Hamiltonian generating complexity on this shared phase space. Concretely, a specific bulk variation —a diffeomorphism $\delta_Y$ dubbed the `new York transformation'— shifts the volume of the maximal slice and, hence, serves as the holographic dual of the boundary complexity shift:
\begin{equation} 
\delta\mathcal{C}_{\text{\tiny CV}}=\Omega_{\text{bulk}}(\delta_Y,\delta) = \Omega_{\text{bdy}}(\delta_Y,\delta)\,. 
\end{equation} 
This construction provides an entry in the holographic dictionary: the maximal volume defines a specific functional of the CFT state (\ref{generalbdydef}), which can be identified as complexity.

Appendix \ref{app:CVforJT} details this mechanism in JT gravity, where the bulk functional $W$ may now include the dilaton field $\Phi$. There we identify a specific bulk perturbation $\delta_Y$ such that the symplectic form pairs it with an arbitrary variation, yielding the change in bulk complexity: $\Omega_{\text{bulk}}(\delta_Y,\delta) = \delta W,$ for any small perturbation $\delta$ of the fields. For CV, in particular, the functional $W$ must equal the volume multiplied by the dilaton $\Phi$, ensuring the proper degrees of freedom are accounted for \cite{Carrasco:2023fcj}. Indeed, in this case we find $\Omega_{\text{bulk}}(\delta_Y,\delta) = \delta(\text{Vol}\cdot\Phi)$. This relation provides the sought-after `first law' for complexity: an infinitesimal deformation of the bulk state yields a proportional change in $W$. 

More generally, in principle, it is possible to construct the appropriate variation for \emph{any} $W$ using the covariant Peierls bracket \cite{Belin:2022xmt}, a systematic method to identify the generator of an observable in Lagrangian field theory. Applied to CAny proposals, it ensures that for any functional $W[g_{\mu\nu}, \Phi,\dots]$, one can define a phase-space vector field $\delta_W$ such that
\[
\Omega_{\text{bulk}}(\delta_W,\delta) = \delta W\,,
\]
for all variations $\delta$ of the fields. In other words, $\delta_W$ is the unique perturbation (up to normalization) that increases $W$ to first order; it is the Hamiltonian vector field of $W$. By bulk/boundary symplectic equivalence, $\delta_W$ also defines a vector in the boundary theory's state space, and the above equation holds for $\Omega_{\text{bdy}}$. This directly informs the boundary dual of complexity: the \emph{increase} in the bulk observable $W$ under $\delta_W$ corresponds to an \emph{increase} in boundary complexity. Practically, if $W$ is the holographic complexity functional, its rate of change along $\delta_W$ equals that of complexity in the CFT. This gives a generalized first law of complexity, relating $\delta \mathcal{C}_{\text{\tiny CAny}}$ to $\delta W$ at the linearized level.

The freedom to choose different $W$ in the bulk translates to a freedom in defining complexity in the CFT. Changing the bulk proposal corresponds to selecting a different `Hamiltonian' (generator) on the common phase space, defining a new direction $\delta_{W'}$ along which complexity growth is measured. Importantly, this does \emph{not} alter the symplectic structure; it amounts to a \emph{canonical transformation} that preserves $\Omega$ while redefining the complexity functional. From the boundary perspective, this reflects the freedom to choose different elementary gates or cost functions in the dual quantum circuit. Complexity is inherently scheme-dependent, relying on the choice of computational basis, gate set, and assigned costs. A unitary basis change can swap ‘simple’ and ‘complex’ operations, altering the numerical complexity of a target state without affecting the state itself.

In the holographic context, two bulk observables $W$ and $W'$ related by a field redefinition or gravitational variable change correspond to two different, but related, complexity measures in the CFT. For instance, one might label certain CFT operators as ‘easy’ to match a bulk volume proposal, whereas a more intricate bulk functional (involving curvature or bulk fields) corresponds to a gate basis where those operators have higher cost while another combination is cheaper. The underlying physics —the trajectory through state space— remains unchanged, but the `benchmark' for measuring its length/complexity shifts. This perspective clarifies why multiple bulk proposals can be valid: they correspond to different choices of basis in the boundary theory’s gate definitions, i.e., canonical transformations on the dual phase space that redefine the complexity measure.

\begin{figure}[t!]
\centering
\begin{tikzpicture}[scale=.9]
	\pgfmathsetmacro\myunit{4}
	\draw[white]	(0,0)			coordinate (a)
		--++(90:\myunit)	coordinate (b);
	\draw [white] (b) --++(0:\myunit)		coordinate (c)
							node[pos=1, above] {\color{black} $\Omega_{\text{bdy}}(\delta_W,\delta)$};
                            
    \draw[white] (c) --++(-90:\myunit)	coordinate (d);                        
    \draw[line width=0.45mm] (a) -- (d) ;
    \draw[line width=0.45mm] (d) --++(0:\myunit)	coordinate (e);
    \draw[line width=0.45mm] (a) -- (e) node[pos=.5, below] {$\text{complexity=anything}$};
    \draw[line width=0.45mm] (c) -- (e) node[pos=1, below] {$\delta W$}
                                node[pos=.5, above, sloped] {$\text{Linearized EOM}$};
    \draw[line width=0.45mm] (a) -- (c) node[pos=0, below] {$\delta \mathcal{C}_{\text{\tiny CAny}}$}
                        node[pos=.5, above, sloped] {$\text{Boundary first law}$};
\end{tikzpicture}
\caption{Trinity of holographic complexity \cite{Carrasco:2023fcj} upgraded to include the full landscape of complexity measures: bulk complexity=anything, $\delta\mathcal{C}_{\text{\tiny CAny}}=\delta W$, boundary first law $\delta\mathcal{C}_{\text{\tiny CAny}}=\Omega_{\text{bdy}}(\delta_W,\delta)$, and bulk-boundary symplectic equivalence for on-shell deformations.
}
\label{fig:trinity}  
\end{figure}
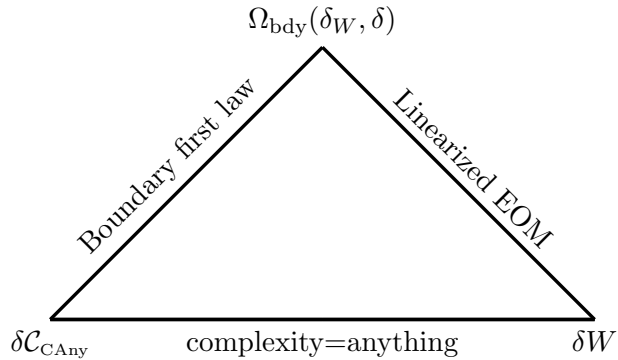

In summary, complexity=anything provides a \emph{dictionary} of complexity measures between bulk and boundary. Once a bulk functional $W$ is chosen (e.g., volume, action, or a higher-curvature invariant), the dual CFT must have a matching complexity functional, ensuring that for any small perturbation, complexity varies as $\delta W$. Different $W$ choices yield distinct first laws of complexity, e.g., $\delta \mathcal{C}_{\rm \tiny CV} \sim \delta(\text{Vol})$ vs. $\delta \mathcal{C}_{\rm \tiny CA} \sim \delta(I_{\text{WdW}})$, each defining a different generator on phase space. Despite these variations, all definitions are physically equivalent up to a change of basis. Enforcing the bulk first law for each $W$ in on-shell perturbations provides a unique notion of boundary complexity. Conversely, consistency between bulk and boundary first laws constrains bulk dynamics, imposing the corresponding Einstein equations (see Fig.~\ref{fig:trinity}). 
Since all such $W$ are related by canonical transformations, no choice is preferred \emph{a priori} —one selects the most convenient definition for a given context.
Holographically, this means complexity in the CFT is not a fixed quantity but a family of observables defined by how operator costs are assigned. Complexity=anything thus enriches the boundary interpretation of holographic complexity, framing it as both a `choice of gauge' (or basis) issue and a dynamical question. While an optimal gate set, cost function, and corresponding bulk dual $W$ may exist for specific scenarios, the deeper insight is that \emph{changing bulk complexity is equivalent to changing the boundary’s computational basis}. This reinforces the view that holographic complexity —like entanglement entropy— has a robust gravitational dual, though inherently scheme-dependent due to the freedom of canonical transformations in its definition.

\section{Conclusions and outlook} \label{sec:disc}

In this article, we developed a general framework for codimension-one measures of holographic complexity in a broad class of two-dimensional dilaton gravity theories. Motivated by the spherical dimensional reduction of codimension-one observables in higher-dimensional gravity, we constructed a more general family of diffeomorphism-invariant functionals tailored to the 2D setting. These observables were shown to exhibit hallmark features of quantum complexity, including late-time linear growth and the switchback effect, thereby satisfying essential criteria expected of any viable complexity proposal. Our results extend the complexity=anything paradigm to the AdS$_2$/CFT$_1$ correspondence and offer a concrete benchmark for studying quantum complexity in low-dimensional holographic dualities. Notably, our proposal for generalized complexity functions made no assumption about behavior or form of the dilaton potential. As such, our proposal applies equally well to 2D dilaton models with AdS$_{2}$ asymptotics, such as deformations of AdS$_{2}$ JT gravity \cite{Witten:2020ert,Witten:2020wvy,Maxfield:2020ale}, or AdS$_{2}$ flow geometries \cite{Anninos:2017hhn,Anninos:2020cwo,Anninos:2022hqo}, or beyond, such as de Sitter JT gravity, which presumably captures the hyperfast complexity growth featured by cosmological horizons \cite{Susskind:2021esx,Jorstad:2022mls}. Pointedly, our proposal holds for all two-dimensional theories of gravity for which complexity exhibits late time linear growth and the switchback effect.

Along the way, we examined the structure of extremal surfaces in detail, offering heuristic interpretations for the appearance of multiple locally extremal solutions—either as distinct circuit realizations of the same final state or as additive contributions from quasi-decoupled sectors. Finally, we offered some comments on the bulk-to-boundary dictionary afforded by the complexity=anything framework, emphasizing its natural embedding within the covariant phase space formalism and the role of the Peierls bracket in identifying consistent boundary duals.

 There are a number of directions to take our work, as we now briefly describe.

\vspace{1mm}

\noindent \textbf{Codimension-zero observables.} While our analysis has focused on codimension-one functionals, a natural extension is to systematically explore the role of codimension-zero observables in 2D gravity. These functionals, which generalize the CA and CV2.0 proposals, involve extremization over bulk regions rather than hypersurfaces and are expected to exhibit similar late-time features, such as linear growth and the switchback effect. In higher-dimensional settings, such observables have been shown to capture complementary aspects of interior dynamics and may offer distinct advantages in encoding quantum information properties of the dual state. Extending the complexity=anything framework to codimension-zero observables in AdS$_2$/CFT$_1$ could thus enrich the landscape of complexity measures, particularly in capturing contributions from bulk matter or topological terms.

\vspace{1mm}

\noindent \textbf{Precise bulk-to-boundary map.} In Section~\ref{sec:dictionary}, we discussed how Peierls' formalism enables the definition of a particular field-space perturbation, $\delta_W$, associated with a given bulk observable $W$, such that the covariant symplectic form satisfies $\Omega_{\textup{bulk}}(\delta_W, \delta) = \delta W$. From the perspective of covariant phase space, this identifies a vector field $X_W^{\textup{bulk}}$ that generates canonical flow along the direction of increasing complexity. By the equivalence of symplectic structures in holography, this bulk vector must correspond to a boundary vector field $X_W^{\textup{bdry}}$ that governs the evolution of sources in the dual theory’s phase space. If the boundary trajectory defined by $X_W^{\textup{bdry}}$ extremizes the complexity functional in eq. \eqref{generalbdydef}, it may be possible to reconstruct the cost function uniquely associated with $W$. We illustrated this mechanism in Appendix \ref{app:CVforJT} for the case of CV in JT gravity, where we identified the deformation that acts as the canonical generator of dilaton-weighted volume. More broadly, this framework opens a pathway toward systematically mapping bulk complexity proposals to precise boundary definitions—an avenue currently under active investigation \cite{Carrasco:2025}.

\vspace{1mm}

\noindent \textbf{Dual microscopic models.} Here we focused on developing bulk, gravitational observables in 2D gravity that exhibit the essential properties of computational complexity. We have not considered any particular dual microscopic model to compare our complexity=anything proposal to. It would be worth leveraging the connection between certain models of 2D dilaton-gravity and one-dimensional quantum mechanical models to quantitatively test our bulk proposal \cite{Rabinovici:2023yex,Ambrosini:2024sre,Heller:2024ldz}. For example, in \cite{Heller:2024ldz}, the authors found precise  agreement between Krylov spread complexity in the double-scaled SYK model and volume complexity in sine-dilaton gravity. It would be interesting to study the generalized complexities presented here using a sin potential and compare to Krylov complexity. 

\vspace{1mm}

\noindent \textbf{Semi-classical corrections and bulk complexity.} An advantage of working with 2D dilaton theories of gravity is that semi-classical backreaction effects due to quantum matter can be exactly incorporated. In this way, one can in principle incorporate quantum corrections to holographic complexity and see how the observable is modified. Such quantum corrections to complexity=volume were previously considered in the context of the Russo-Thorlacius-Susskind model \cite{Schneiderbauer:2019anh,Schneiderbauer:2020isp}, and in semi-classical JT gravity in \cite{Carrasco:2023fcj}. Notably, in \cite{Carrasco:2023fcj} it was found that the CV functional was to be modified by a `bulk complexity' term (a functional that presumably characterizes the complexity of bulk quantum matter fields), in order for the first law of complexity to recover the semi-classical gravitational equations in the bulk. It would be relatively straightforward and interesting to extend our work to semi-classical models of dilaton gravity (cf. \cite{Pedraza:2021cvx}). 

\vspace{1mm}

\noindent \textbf{Singularity imprints in 2D.} One of the most tantalizing features of generalized complexities is their ability to access deep regions of the black hole interior, potentially offering geometric insight into singularities. Although 2D gravities do not contain curvature singularities, one can still track the imprints of higher-dimensional black hole singularities upon dimensional reduction. These effects are encoded in the dilaton profile, which typically diverges where the parent theory harbors a singularity. In the context of holographic complexity, certain proposals can detect the presence of singularities \cite{Jorstad:2023kmq}, while more refined diagnostics—such as the extraction of Kasner exponents near generic BKL-type singularities—are accessible via the approaches of \cite{Arean:2024pzo,Caceres:2024edr}. It would be interesting to explore these effects from a purely 2D perspective, perhaps via dimensional reduction of near-extremal charged black holes with Kasner interiors, such as those constructed in \cite{Hartnoll:2020rwq,Hartnoll:2020fhc,Henneaux:2022ijt,Carballo:2024hem}.

\vspace{1mm}

We hope to come back to these interesting questions in the near future.

\section*{Acknowledgments}

We would like to thank José Barbón, Francesco Gentile, Carlos Perez-Pardavila and Shan-Ming Ruan for helpful discussions and comments on the manuscript. EC and VP are supported by the National Science Foundation under grant PHY–2210562. EC thanks the Instituto de F\'isica Te\'orica (IFT) UAM/CSIC, Madrid, for its hospitality and acknowledges the support of the IFT Severo Ochoa Associate Researcher program. RC and JFP are supported by the ‘Atracci\'on de Talento’ program (Comunidad de Madrid) grant 2020-T1/TIC-20495, the Spanish Research Agency via grants FPU22/01262, CEX2020-001007-S and PID2021-123017NB-I00, funded by MCIN/AEI/10.13039/501100011033, and by ERDF `A way of making Europe.' RC also acknowledges support from the CSIC iMOVE mobility program, fellowship 24101, and thanks the Theory Group at The University of Texas at Austin for its hospitality during the final stage of this project. AS is supported by the STFC consolidated grant ST/X000753/1, and partially funded by the Royal
Society under the grant ``Concrete Calculables in Quantum de Sitter''.

\appendix 

\section{Dimensional reduction of curvature invariants} \label{app:dimredx}

Here we list the dimensionally reduced scalar curvature invariants, under a spherical dimensional reduction. 
Specifically, we start with the spherically symmetric ansatz for a four-dimensional spacetime $\hat{g}_{MN}$, 
\begin{equation}
    ds^2 =\hat{g}_{MN}dX^{M}dX^{N}= g_{\mu\nu}(x) dx^{\mu}dx^{\nu} + \ell^2\Phi(x)d\Omega_{2}^2\;, 
\end{equation}
where $\ell$ is some four-dimensional length scale, and $\Phi$ is the dilaton. As described in the main text (see also Appendix A of \cite{Svesko:2022txo}), we further perform the following Weyl transformation on the two-dimensional metric $g_{\mu\nu}$
\begin{equation}
\begin{split}
    g_{\mu\nu} &\rightarrow \omega^2 g_{\mu\nu}\;,\quad \omega= \gamma\Phi^{-1/4} 
\end{split}
\end{equation}
to eliminate kinetic terms for the dilaton $\Phi$ that appear in the dimensionally reduced Einstein-Hilbert action. Here $\gamma$ is some unspecified real constant that can be chosen for convenience.  With the aid of Mathematica package xTensor 
we then compute the dimensionally reduced curvature invariants  $\hat{R}$, $\hat{R}_{MN}^{2}$, and $\hat{R}_{MNPQ}^{2}$ for the four dimensional spacetime:
\begin{equation}
    \hat{R} = \frac{2 (\frac{2 \gamma^2}{\Phi^{1/2}} + \ell^2 R \Phi) - 3 \ell^2 \
\Box\Phi}{2 \ell^2 \gamma^2 \Phi^{1/2}}\;,
\end{equation}
for $\Box\Phi\equiv g^{\mu\nu}\nabla_{\mu}\nabla_{\nu}\Phi$,
\begin{equation}
\begin{split}
    \hat{R}_{MN}\hat{R}^{MN} & = \frac{1}{{64 \ell^4 \gamma^4 \Phi^3}}\Bigg( 64 \ell^4 R_{\mu\nu} R^{\mu\nu} \Phi^4 + \ell^2 \nabla_{\mu}\Phi \
\nabla^{\mu}\Phi (80 \gamma^2 \Phi^{1/2} - 8 \ell^2 R \Phi^2 \\
    & + 25 \ell^2 \
\nabla_{\mu}\Phi \nabla^{\mu}\Phi - 52 \ell^2 \Phi \Box\Phi) + 8 \Phi \
\Bigl(16 \gamma^4 - 6 \ell^4 R_{\mu\nu} \Phi \nabla^{\mu}\Phi \nabla^{\nu}\Phi  \\
    & + 3 \ell^4 \nabla^{\mu}\Phi \nabla^{\nu}\Phi \nabla_{\nu}\nabla_{\mu}\Phi  - 8 \ell^4 R_{\mu\nu} \Phi^2 \
\nabla^{\mu}\nabla^{\nu}\Phi \\
    & +  \bigl(2 \ell^4 \Phi \nabla_{\nu}\nabla_{\mu}\Phi \
\nabla^{\nu}\nabla^{\mu}\Phi + \Box\Phi (-16 \ell^2 \gamma^2 \Phi^{1/2}  + 4 \
\ell^4 R \Phi^2 + 3 \ell^4  \Phi \Box\Phi)\bigr)\Bigr)\Bigg)\;,\\
\end{split}
\end{equation}

\begin{equation}
\begin{split}
    \hat{R}_{MNPQ}\hat{R}^{MNPQ}=& \frac{1}{128 \ell^4 \gamma^8 \Phi^2}\Biggl( 128 \ell^4 R_{\mu\nu\rho\sigma} R^{\mu\nu\rho\sigma} \Phi^4 + \ell^2 \nabla_{\mu}\Phi \
\nabla^{\mu}\Phi \bigl(177 \ell^2 \nabla_{\nu}\Phi \nabla^{\nu}\Phi \\
& - 8 (8 \gamma^2 \
\Phi^{1/2} + 4 \ell^2 R \Phi^2 + 39 \ell^2 \Phi \Box \Phi)\bigr) \\
& + 8 \Phi (64 \gamma^4 + 48 \ell^4 R_{\mu\nu} \
\Phi^2 \nabla^{\nu}\nabla^{\mu}\Phi - 36 \ell^4 R_{\mu\nu} \Phi \nabla^{\mu}\Phi \nabla^{\nu}\Phi \\
& + 3 \ell^4 \nabla^{\mu}\Phi \nabla^{\nu}\Phi \nabla_{\nu}\nabla_{\mu}\Phi  + 22 \ell^4  \Phi \nabla_{\mu}\nabla_{\nu}\Phi \nabla^{\mu}\nabla^{\nu}\Phi + 8 \ell^4 
\Phi (\Box\Phi)^{2}\Biggr)\;.
\end{split}
\end{equation}


\section{Complexity=volume for JT gravity and the Peierls construction} \label{app:CVforJT}

In \cite{Carrasco:2023fcj}, the analog of the `new York' deformation \cite{Belin:2018bpg} $\delta_{\text{Y}}$ was uncovered for AdS-JT gravity such that the bulk symplectic form 
was proportional to the linear variation of the generalized volume, $\Omega_{\text{bulk}}(\psi,\delta_{Y}\psi,\delta\psi)=\delta W$, for dynamical fields $\phi=\{g_{\mu\nu},\phi\}$ and  generalized volume functional 
\begin{equation}\label{eq:WfuncJT}
    W=\int_{\Sigma} dy \sqrt{h} (\phi+\phi_0).
\end{equation}
In this appendix we will determine the specific perturbations $\delta_W$ to the field content of AdS-JT gravity using the covariant Peierls' construction. An analogous treatment was given for the new York deformation for Einstein gravity and complexity=volume in \cite{Belin:2022xmt}.  
Before we describe the Peierls construction, let us briefly recall some relevant facts about the covariant phase construction of JT gravity (for more details see Section 4.1 of \cite{Carrasco:2023fcj}). 

\subsection*{New York deformation of JT gravity}

AdS-JT gravity in Lorentzian signature has the action
\begin{equation}
   S_0=\int_{\mathcal{M}}d^2 x\sqrt{-g}\left[\phi_0 R +\phi(R+2)\right]\;,
\label{eq:JTactions0}\end{equation}
where we ignore the Gibbons-Hawking-York boundary term needed for the variational problem to well-posed. Varying the action
yields 
\beq \delta S_{0}=\int_{\mathcal{M}}E_{\phi}\delta\phi+\int_{\Sigma}\theta(\psi,\delta\psi)\;,\eeq
where $\psi=\{h_{ab},\phi\}$ and $E_{\psi}$ represents the equation of motion form for each field (with an implicit sum over field type $\phi$), and $\theta$ is the symplectic potential 1-form. 
Performing an ADM split, the symplectic form $\Omega_{\text{bulk}}$ is easily read off in terms of phase space variables $(h_{ab},\pi^{ab},\phi,\pi_{\phi})$
\beq \Omega_{\text{bulk}}(\psi,\delta_{1}\psi,\delta_{2}\psi)=\int_{\Sigma_{t}}(\delta_{1}\pi^{ab}\delta_{2}h_{ab}-\delta_{2}\pi^{ab}\delta_{1}h_{ab})+\int_{\Sigma_{t}}(\delta_{1}\pi_{\phi}\delta_{2}\phi-\delta_{2}\pi_{\phi}\delta_{1}\phi)\;.\label{eq:bulksympformclassJT}\eeq
Here $h_{ab}$ denotes the induced metric of codimension-1 timeslice $\Sigma_{t}$ in the ADM split with conjugate momenta $\pi^{ab}$, and $\pi_{\phi}$ is the conjugate momenta to the dynamical dilaton.

It is easy to verify the deformation which achieves
$\Omega_{\text{bulk}}(\psi,\delta_{Y}\psi,\delta\psi)=\delta W$ for volume (\ref{eq:WfuncJT}) must satisfy
\beq
\delta_Y \pi^{ab} = \frac{\rho}{2}\sqrt{h} h^{ab} (\phi + \phi_0), \quad \delta_Y \pi_{\phi} = \rho \sqrt{h},\quad \delta_{Y}h_{ab}=\delta_{Y}\phi=0\;.
\label{eq:nuNYtransJT}\eeq
for constant $\rho$, such that
\begin{equation}\label{eq:JTVol}
     \Omega_{\text{bulk}}(\psi,\delta_Y \psi, \delta \psi) = \rho \delta W = \rho \int_{\Sigma_{t}}dy[(\delta{\sqrt{h}})(\phi + \phi_0) + \sqrt{h} \delta \phi].
\end{equation}
In terms of configuration space variables the new York deformation behaves as 
\beq \label{eq: configvariables}
\delta_Y K = - \frac{\rho}{2},  \quad  \delta_Y \dot{\phi} = - \frac {\rho N} {2}(\phi+ \phi_0),
\eeq
where $K$ is the trace of the extrinsic curvature for the surface $\Sigma$, and $N$ is the lapse function.
Note that the deformation preserves the constraints of the theory \cite{Carrasco:2023fcj}. 

The goal of this appendix is to directly derive the deformation (\ref{eq: configvariables}) using the Peierls construction.

\subsection*{Peierls bracket: general comments and strategy}

As neatly summarized in \cite{Harlow:2019yfa,Belin:2022xmt}, the Peierls bracket provides an intuitive understanding to the Poisson bracket in covariant phase space. To this end, recall covariant (pre-) phase space can be defined as the  set of solutions of the equations of motion. The Poisson bracket between two functions of phase space $f$ and $g$ is given by 
\beq \{f,g\}\equiv\Omega(X_{g},X_{f})=X_{g}\delta f=-X_{f} \delta g\;,\eeq
for symplectic form $\Omega$ and Hamiltonian vector fields $X_{f}$ and $X_{g}$ conjugate to $f$ and $g$, respectively. 
The right hand side characterizes the (covariant) Peierls bracket.\footnote{The definition of the Peierls bracket does not involve the symplectic form, in fact. Rather, the definition is \cite{Harlow:2021dfp}: $\{f,g\}[\psi]\equiv\partial_{\rho}f[\psi+\rho\delta_{g}\psi]|_{\rho=0}$ for solution $\psi$ to the equations of motion and $\delta_{g}\psi$ is a particular solution to the linearized equations of motion about $\psi$.} To understand Hamiltonian evolution on the phase space covariantly one must thus construct the appropriate function $X_{g}$ (or $X_{f}$), for which Peierls developed a prescription \cite{Peierls:1952cb}.\footnote{Recall that in ordinary classical mechanics, the usual recipe to compute the Poisson bracket requires one introduce a notion of time to make a coordinate transformation between generalized coordinates $\{q\}$ and their velocities $\{\dot{q}\}$ to phase variables. This destroys manifest Lorentz covariance. The Peierls bracket is manifestly covariant.} 

The algorithm is as follows (see \cite{Harlow:2019yfa} for more details). Take $g$ to be a function on the configuration space which is constructed using  dynamical fields $\psi^{i}$ over a finite window bounded by past/future Cauchy slices $\Sigma_{\pm}$. Then, 

\vspace{2mm}

\noindent (1) Introduce a deformed action $S\equiv S_{0}-\rho g$, for small parameter $\rho$. The equations of motion of $S$ differ from the unperturbed theory $S_{0}$ in the region bounded by $\Sigma_{\pm}$. The deformed action will be stationary about field configurations obeying the deformed equations of motion $E_{i}-\rho\Delta^{g}_{i}=0$ where $\Delta^{g}_{i}$ in general is some spacetime $d$-form which vanish outside the region bounded by $\Sigma_{\pm}$ and obey $\delta g=\int_{\mathcal{M}}\Delta^{g}_{i}\delta\psi^{i}$.  

\vspace{2mm}

\noindent (2) Pick a solution $\psi^{i}_{0}$ to the unperturbed equations of motion. To linear order, the solution to the perturbed equations of motion takes the form $\psi^{i}=\psi_{0}^{i}+\rho \mathfrak{h}^{i}$, where $\mathfrak{h}^{i}$ is a solution to the deformed equations of motion linearized about a solution to the unperturbed equations of motion; we write $\mathfrak{h}^{i}=\delta\psi^{i}$. Two particular solutions of relevance include the advanced solution $\mathfrak{h}^{i}_{A}\equiv\delta_{A}\psi^{i}$, which obeys $\mathfrak{h}^{i}_{A}|_{J^{+}(\Sigma_{+})}=0$, for causal future $J^{+}(\Sigma_{+})$. Similarly, there is the retarded solution $\mathfrak{h}_{R}^{i}\equiv\delta_{R}\psi^{i}$, which vanishes in the causal past of $\Sigma_{-}$.

\vspace{2mm}

\noindent (3) The appropriate Hamiltonian vector field is then $X_{g}\equiv X^{\{\mathfrak{h}_{R}\}}-X^{\{\mathfrak{h}_{A}\}}$, for configuration-space vector field $X^{\{\mathfrak{h}\}}\equiv\int d^{2}x \mathfrak{h}^{i}\frac{\delta}{\delta\psi^{i}}$ (which generally obeys $X^{\{\mathfrak{h}\}}\delta E_{i}=\Delta^{g}_{i}$). Explicitly, $X_{g}\equiv \int d^{2}x(\delta_{R}\psi^{i}-\delta_{A}\psi^{i})\frac{\delta}{\delta\psi^{i}}$. 

\vspace{2mm}

From the final step, it is easy to see what the conjugate variation $\delta_{w}\psi$ must be such that $\Omega(\delta,\delta_{W})=\delta W$. Specifically, modifying step (1) such that the perturbed action is $S\equiv S_{0}-\rho W$ for diffeomorphism invariant functional $W$. Then, 
\beq \delta_{W}\psi\equiv \delta_{R}\psi-\delta_{A}\psi\;.\label{eq:conjugatevariation}\eeq
Below we will follow the strategy outlined above to determine the appropriate $\delta_{W}$ such that $\Omega_{\text{bulk}}(\delta,\delta_{W})=\delta W$ for JT gravity with volume functional (\ref{eq:WfuncJT}).

\subsection*{Conjugate variation for JT gravity}

Here we closely follow \cite{Harlow:2021dfp} (see their Section 4.1). We follow Peierls prescription, and perturb the JT action (\ref{eq:JTactions0}) via the volume functional $W$ \eqref{eq:WfuncJT}.
The next step consists on finding the equations of motion to this new action $S=S_0-\rho W$. Notice that the variation of the first piece in this action will be equal to the unperturbed equations of motion of JT gravity theory, given by \eqref{eq:dilaton_theory_eom} after taking $V(\Phi)=2\Phi$. Thus, we only have to focus on finding the variation of the perturbation $W$.

First, it is useful to recast $W$ as follows. 
Note that the surface $\Sigma$ appearing in \eqref{eq:WfuncJT} is a one-dimensional curve, which for convenience we describe in terms of the pseudo-affine parameter $\lambda$ 
obeying
\begin{equation}
    \mathfrak{L} d\lambda=\sqrt{dy^\mu dy^\nu g_{\mu\nu}}\Rightarrow \mathfrak{L}=\sqrt{\frac{dy^\mu}{d\lambda}\frac{dy^\nu}{d\lambda}g_{\mu\nu}}=\textup{const,}
\end{equation}
where $y^\mu$ is the embedding of the curve into the two-dimensional space and $\mathfrak{L}$ is the length of $\Sigma$. Then
\begin{equation}
    W=\int_0^1 d\lambda(\phi+\phi_0)\sqrt{\frac{dy^\mu}{d\lambda}\frac{dy^\nu}{d\lambda}g_{\mu\nu}}\;.
\end{equation}
Its linear variation is
\begin{equation}
\begin{split}
    \delta W=&\int_0^1 d\lambda\left[\sqrt{\dot{y}^{\mu}\dot{y}^{\nu}g_{\mu\nu}}\delta \phi+\frac{1}{2}(\phi+\phi_0)\frac{\dot{y}^\mu \dot{y}^\nu\delta g_{\mu\nu}}{\mathfrak{L}}\right]\;,
\end{split}
\label{eq:deltaWJT}\end{equation}
with $\dot{y}^\mu\equiv\frac{dy^\mu}{d\lambda}$ being tangent to the curve $y^{\mu}$.

It is useful to further recast the two terms in $\delta W$. Focusing on the first term, we write
\begin{equation}
    \begin{split}
    \int_0^1 d\lambda\sqrt{\dot{y}^\mu\dot{y}^\nu g_{\mu\nu}}\delta \phi=\mathfrak{L}\int_0^1 d\lambda \delta \phi =\mathfrak{L} \int_{\mathcal{M}} d^2 x \sqrt{-g}\delta{\phi}\int_{0}^1 d\lambda \delta^2(x-y(\lambda)).
    \end{split}
\end{equation}
Introduce Gaussian normal coordinates, where the metric reads 
\begin{equation}\label{eq:metric}
    ds^2 =-dn^2+\ell(n,\lambda)d\lambda^2\;,
\end{equation}
with $\ell(n,\lambda)$ being the induced metric on the surfaces of constant $n$. Let $\Sigma$ is be the surface located at $n=0$, such that $\ell(0,\lambda)=\mathfrak{L}$. Consequently, 
\begin{equation}
    \int_0^1 d\lambda\sqrt{\dot{y}^\mu\dot{y}^\nu g_{\mu\nu}}\delta \phi=\int_{\mathcal{M}} d^2 x \sqrt{-g}\delta{\phi}\delta(n).
\end{equation}
Now, we proceed with the second term in \eqref{eq:deltaWJT}: 
\begin{equation}
\begin{split}
    &\frac{\mathfrak{L}}{2}\int_0^1 d\lambda(\phi+\phi_0)\frac{\dot{y}^\mu \dot{y}^\nu\delta g_{\mu\nu}}{\mathfrak{L}^2}=\frac{\mathfrak{L}}{2}\int_0^1 d\lambda(\phi+\phi_0)e^\mu e^\nu\delta g_{\mu\nu}\\
    &=\frac{\mathfrak{L}}{2}\int_\mathcal{M} d^2x\sqrt{-g}(\phi+\phi_0)e^\mu(x)e^\nu(x)\delta g_{\mu\nu} \int_0^1\delta \lambda'  \delta^2 (x-y(\lambda'))\\
    &=\frac{1}{2}\int_\mathcal{M} d^2 x\sqrt{-g}(\phi+\phi_0)\delta(n) e^\mu e^\nu \delta g_{\mu\nu}\,,
\end{split}
\end{equation}
for unit tangent vector $e^\mu\equiv\frac{1}{\mathfrak{L}}\frac{dy^\mu}{d\lambda}$.  Combined, the variation of the observable $W$ is
\begin{equation}
    \delta W=\int_{\mathcal{M}} d^2 x \sqrt{-g}\left[\delta{\phi}\delta(n)+\frac{1}{2}(\phi+\phi_0)\delta(\tau) e^\mu e^\nu \delta g_{\mu\nu}\right]
\end{equation}
Together with the variation of the unperturbed action $S_0$ yields the perturbed equations of motion
\begin{equation}\label{eq: perturbed eom}
    \begin{split}
    R+2&=\rho\delta(n)\\
    \nabla^\mu\nabla^\nu \phi +g^{\mu\nu}(\phi-\nabla^2\phi)&=\frac{\rho}{2}(\phi+\phi_0)\delta(n)e^\mu e^\nu
    \end{split}
\end{equation}
When $\rho=0$, one recovers the equations of motion of the unperturbed theory $S_{0}$. 

The next step in the Peierls construction is to determine the advanced and retarded solutions $\psi_{0}+\rho\delta_{R,A}\psi$. For our purposes, we are only interested in the difference $\delta_{R}\psi-\delta_{A}\psi$. We can in fact find this difference without having to explicitly solve for the advanced/retarded solutions. Rather, it is enough to see how 
the addition of the term $W$ in the action introduces a jump in the fields. Adapting the spirit of \cite{Belin:2022xmt}, integrate the left-hand side of the first perturbed equation of motion (\ref{eq: perturbed eom}) across $n=0$ 
\begin{equation}
    \int_{-\epsilon}^\epsilon dn (R+2)=-2\int_{-\epsilon}^\epsilon \nabla_\mu(n^\nu\nabla_\nu n^\mu-n^\mu K)dn = -2\Delta K\;,
\end{equation}
where $\epsilon$ is some small parameter, and we implemented the Gauss-Codazzi equations to express the two-dimensional Ricci scalar in terms of the curvatures of $\Sigma$.\footnote{The relevant Gauss-Codazzi equations are $R=R^{(1)}-(K^2-K_{ab}^2)-2\nabla_\mu(n^\nu\nabla_\nu n^\mu-n^\mu K)$, where $R^{(1)}$ is the Ricci scalar of $\Sigma$, $K$ is its extrinsic curvature tensor and $n^\mu$ is its normal vector. Here $R^{(1)}=0$.} Integrating the right-hand side of the (\ref{eq: perturbed eom}) then gives 
\begin{equation}
    \Delta K=-\frac{\rho}{2}.
\end{equation}
The jump discontinuity in $K$ is controlled by the deformation parameter $\rho$. 

Let us proceed and analyze the dilaton equation of motion to the perturbed theory. Contracting with $e_\mu e_\nu$ and using  $\nabla^2 \phi=(e^\mu e^\nu-n^\mu n^\nu)\nabla_\mu\nabla_\nu \phi$ gives
\begin{equation}\label{eq:junction condition phi}
    \ddot{\phi}+\phi=\frac{\rho}{2}(\phi+\phi_0)\delta(n).
\end{equation}
 Again, integrate  across $n=0$
\begin{equation}
    \Delta \dot{\phi}+\int_{-\epsilon}^\epsilon \phi dn =-\frac{\rho}{2}(\phi+\phi_0)|_{n=0},
\end{equation}
Like the metric, the dilaton is taken to be continuous across the surface at $n=0$, such that the second term vanishes. We write $\Delta\phi$. There is, however, a jump discontinuity in the `time' derivative of $\phi$, given by
\beq
\Delta \dot{\phi}=-\frac{\rho}{2}(\phi+\phi_0)|_{n=0}\;.
\eeq

Finally, 
near the surface $\Sigma$, the retarded solution looks like $\delta_R\psi= -\Theta(n)\Delta \psi$, while the advanced solution looks $\delta_A \psi=\Theta(-n)\Delta \psi$. Their difference gives $\delta_W \psi=\delta_R\psi-\delta_A \varphi=-[\Theta(n)+\Theta(-n)]\Delta \varphi=-\Delta \varphi$. This fact implies that the discontinuities above may be identified with the variations \cite{Belin:2022xmt}
\begin{equation}
    \delta_W K=\frac{\rho}{2},\quad \delta_W \dot{\phi}=\frac{\rho}{2}(\phi+\phi_0)|_{n=0}.
\end{equation}
These variations coincide with (\ref{eq: configvariables}) (cf. Eq. (4.22) in \cite{Carrasco:2023fcj}).

\bibliographystyle{JHEP}
\bibliography{refs.bib}
\end{document}